\documentclass[12pt]{article}
\usepackage{amsmath,amsfonts,amsthm,bm} 
\usepackage{graphicx,psfrag,epsf}
\usepackage{enumerate}
\usepackage{natbib}
\usepackage{mathrsfs}
\usepackage{url}
\usepackage{chngcntr}
\usepackage{dsfont }
\usepackage[dvipsnames]{xcolor}
\usepackage{relsize}
\usepackage{amssymb }
\usepackage{enumitem}
\usepackage[english]{babel}	
\usepackage{upgreek }
\usepackage{multirow}
\usepackage[toc,page]{appendix}
\usepackage{boldline}
\usepackage{caption, subcaption}
\usepackage{ulem}
\usepackage{graphicx}
\DeclareGraphicsExtensions{.pdf,.jpeg,.png}
\usepackage{epstopdf}
\usepackage[colorlinks=TRUE, citecolor=blue,linkcolor=blue,urlcolor=blue]{hyperref}
\newtheorem{theorem}{Theorem}
\newtheorem{remark}{Remark}

\newtheorem{corollary}{Corollary}

\newcommand{\ve}[1]{\mbox{\boldmath ${#1}$}}

\usepackage{authblk}

\def\T{{ \mathrm{\scriptscriptstyle T} }}

  \title{Joint Curve Registration and Classification with Two-level Functional Models}
   \author[1]{Lin Tang \thanks{The first two authors have equal contributions}} 
  \author[2]{Pengcheng Zeng}
  \author[3]{Jian Qing Shi \thanks{Corresponding author, email: j.q.shi@ncl.ac.uk}}
  \author[4]{Won-Seok Kim}

  \affil[1]{School of Mathematics and Statistics, Yunnan University, China}
  \affil[2]{Department of Statistics, The Chinese University of Hongkong, HK, SAR China}
  \affil[3]{School of Mathematics, Statistics and Physics, Newcastle University, UK}
  \affil[4]{Department of Rehabilitation Medicine, Seoul National University College of Medicine, Seoul National University Bundang Hospital, Seongnam, South Korea}
 
  \date{\today}

\begin{document}
  \maketitle
  	
\bigskip
\begin{abstract}
Many classification techniques when the data are curves or functions have been recently proposed. However, the presence of misaligned problems in the curves can influence the performance of most of them. In this paper, we propose a model-based approach for simultaneous curve registration and classification. The method is proposed to perform curve classification based on a functional logistic regression model that relies on both scalar variables and functional variables, and to align curves simultaneously via a data registration model. EM-based algorithms are developed to perform maximum likelihood inference of the proposed models. We establish the identifiability results for curve registration model and investigate the asymptotic properties of the proposed estimation procedures. Simulation studies are conducted to demonstrate the finite sample performance of the proposed models. An application of the hyoid bone movement data from stroke patients reveals the effectiveness of the new models.

\end{abstract}

\noindent%

\section{Introduction}
\label{sec:intro}

Data in various fields such as biomedicine, econometrics and epidemiology are collected in the form of curves, so it is natural to consider curves as observations and covariates  \citep{shi2012}. Methodologies focusing on the curves themselves, as the objects of interest, is termed as Functional Data Analysis (FDA) \citep{ram05}. Recently, FDA has received increasing attention in many areas, please see, for example, \cite{ullah2013}, \cite{yu2016} and \cite{mallor2018}, among others. In particular, functional data classification, which consists of identifying predefined discrete class labels based on the observed curves or images, has been fruitfully exploited. For example, \cite{james2002} and \cite{muller2005} extended generalized
functional linear models to the case of sparse and irregular data, with an emphasis on functional binary regression and functional logistic discrimination for the classification of longitudinal data; \cite{james2001} developed the linear discriminant analysis (LDA) to functional data classification; \cite{li2008} proposed the so called functional segment discriminant analysis (FSDA), which combines the LDA and the support vector machine as the classifier; \cite{zhu2012} proposed to use mixed models to perform classification and feature extraction on functional data. Other related work include \cite{delaigle2012achieving}, \cite{mosler2014fast} and \cite{chamroukhi2019model}.

In many situations, however, such as the motion analysis of the hyoid bone of stroke patients \citep{kim2017} discussed in Sect.~\ref{sec:4}, a specific aspect of the curve data is the presence of misaligned problems. When the trajectories of the hyoid bone movement during swallowing are recorded, peaks and valleys for phases of the trajectories include elevation, anterior, remain and return occur at different times for different individuals. A large fraction of the variability in a sample of trajectories is then best explained as time variation \citep{mullerH2005}. In this context, the corresponding warping (curve registration, alignment) procedures can be used, for example, landmark method \citep{gasser1995}, procrustes method \citep{ramsay1998}, time acceleration models \citep{capra1997}, maximum likelihood based alignment \citep{ronn2001}, time synchronization of random processes \citep{liu2004} and some other forms of curve registration \citep{wang1997, kneipetal2000}. However, most of the existing methods are performed as preprocessing steps prior to the final statistical analysis of the misaligned curves. Treating curve registration as a preprocessing step may cause problems. In particular, the noise-corrupted observations from collection of trajectories of the hyoid bone movement can skew registration results such that noise rather than signal is aligned. Moreover, in the motion analysis of hyoid bone, the knowledge of curve class labels provides valuable information in predicting an individual's transformed trajectory.

To address the above mentioned issues, limited work has been done to infer functional clusters and register curves simultaneously. For example, \cite{liu2009simultaneous} developed a coherent clustering procedure that allows for simultaneously aligning and clustering k-centres functional data; \cite{zhang2014joint} proposed a hierarchical model combines a Dirichlet process mixture model for clustering of common shapes of curves with a reproducing kernel representation of phase variability for registration; \cite{sangalli2010k-mean} proposed a K-mean alignment procedure for jointly clustering and aligning curves. However, the above models are not applicable to multi-dimensional curve observations and did not take subject-specific information into consideration. To solve this problem, more recently, \cite{zeng2019} proposed a simultaneous registration and clustering model which combines a Gaussian process functional regression model with time warping and an allocation model depending on both scalar and functional predictors. However, \cite{zeng2019} did not give any theoretical properties of their proposed estimates.

In the light of the above, we propose a joint modelling procedure for simultaneous curve registration and classification. We prove the identifiability results of the proposed data registration model under some mild conditions. In addition, the asymptotic properties of the proposed estimation procedures are also investigated. Simulation studies and a collection of trajectories of hyoid bone movement obtained from stroke patients are used to demonstrate the finite sample performance of the proposed models.

We organize this paper as follows. In Sect. \ref{sec:2}, we introduce the two-level functional model for simultaneous curve registration and classification. Model specification and the corresponding estimation procedures are given in Sect. \ref{sec:2.1} and Sect. \ref{sec:2.2}, respectively. We discuss curve prediction approach in Sect. \ref{sec:2.3}. The theoretical properties of the proposed models are given in Sect. \ref{sec:2.4}. In Sect. \ref{sec:3}, simulation studies are conducted. In Sect. \ref{sec:4}, we apply the methodology to motion analysis of hyoid bone of stroke patients. We provide discussion and further development in Sect. \ref{sec:5} and defer the technical conditions and proofs in the online supplemental material.

\section{The joint curve registration and classification models}
\label{sec:2}
\subsection{Model specification}
\label{sec:2.1}
We start by assuming that data are collected across $N$ individuals from two groups labelled as $k=0$ and $k=1$. Let $\ve{x}_{ki}(t_{ij})$ be the two-dimensional curve of the $i$-th subject for $i=1,\ldots,N_k$, where $N=N_0+N_1$ and $N_k$ is the number of individuals in the $k$-th group. Suppose that a number $n_{ki}$ of observations are obtained for the $i$-th subject in the $k$-th group. Then the data set we considered can be denoted as
\[
{\cal D}=\big\{(y_{ki},x_{1ki}(t_{ij}), x_{2ki}(t_{ij}), \ve{v}_{ki}): i=1,\ldots, N_k; j=1,\ldots,n_{ki}; k=0,1\big\},
\]
where $x_{1ki}(t_{ij})$ and $x_{2ki}(t_{ij})$ represent $x$-coordinates and $y$-coordinates of $\ve{x}_{ki}(t_{ij})$, respectively. $y_{ki}$ is a binary outcome represents the class label of the curve and $y_{ki}=k$. $\ve{v}_{ki}$ is the scalar variable, providing information for each specific subject such as patient's age and gender.

We now consider a joint modelling approach for simultaneously aligning and classifying the observed curves. To do so, we propose to use a two-level functional regression model with mixed functional and scalar variables. In the first level, the continuous curves across different subjects in the $k$-th group are modelled as follows
\begin{equation}\label{eq:regist}
x_{aki}(t) = (\tau_{ak}\circ g_{ki})(t) + r_{aki}(t) + \epsilon_{ai}(t), i = 1,\dots,N_{k}; k=0,1; a=1,2,
\end{equation}
where the notation ``$\circ$" denotes functional composition, i.e. $(\tau \circ g)(\cdot) = \tau(g(\cdot))$. $g_{ki}(t)$ is the inverse of a warping function, $\tau_{ak}(\cdot)$ is a fixed but unknown nonlinear mean curve, which without loss of generality, can be modelled by  $\tau_{ak}(t) = \zeta_{a}(t) + \xi_{ak}(t)$, where $\zeta_{a}$ is the underlying profile shared across two groups and $\xi_{ak}$ is the group-specific variation centered around $\zeta_{a}$. Both $\zeta_{a}$ and $\xi_{ak}$ can be approximated by a set of basis functions, the details will be given in the next subsection.  The variation among different subjects is modelled by a non-linear functional random effects, $ r_{aki}(t)$, by a Gaussian process with zero-mean and covariance function $\ve{S}$ \citep{shi2012} parameterized by $\ve{\rho}_s$. $\epsilon_{ai}$ is independent identically distributed Gaussian noise with variance $\sigma^{2}$.

Taking the common effect of the warping functions in the same group and the variation among different subjects into consideration, we model the warping function as follows
\begin{equation}\label{eq:warp}
g_{ki}(t) = t + w_{k}(t) + w_{ki}(t),
\end{equation}
where $w_{k}(\cdot)$ is the fixed part representing the common effect within group $k$, and $w_{ki}(\cdot)$ is the random effects in terms of different subjects. Instead of  making assumptions for $w_{k}$ and $w_{ki}$, we first discretize them by $n_w$ equidistant anchor points in $[0,1]$ \citep{raket2016separating, zeng2019}. Then we model $w_{ki}$ by a zero-mean Gaussian process with covariance function $\ve{H}$ parameterized by $\ve{\rho}_h$. Some other forms of the warping functions can also be used, see \cite{liu2009simultaneous} and \cite{sangalli2010k-mean} for example.

In the second level, a functional logistic regression model with mixed functional and scalar variables is defined as
\begin{equation}\label{eq:response}
y_{i}|\ve{v}_{i}, \ve{x}_{i}(t) \sim {\rm Bernoulli}(\pi_{i}),
\end{equation}
where $\pi_{i} = {\rm Pr}(y_{i} = 1 |\ve{v}_{i}, \ve{x}_{i}(t))$ is specified by the following logistic regression model
\begin{equation}\label{eq:logist}
{\rm logit}(\pi_{i}) = b_{0} + \ve{v}_{i}^{\T}\ve{b_{1}} +\int \ve{x}_{i}(g^{-1}_i(t))\ve{\beta}(t)dt, \hspace{5mm}i = 1,\dots,N,
\end{equation}
where $(b_{0}, \ve{b}^{\T}_{1})^{\T}$ is $(p+1)\times 1$ dimensional scalar coefficients, $\ve{\beta}(t) = (\beta_{1}(t), \beta_{2}(t))^{\T}$ is the functional coefficient.
$\ve{x}_{i}(g^{-1}_i(t)) = \big(x_{1i}(g^{-1}_i(t)), x_{2i}(g^{-1}_i(t))\big)^{\T}$ is the aligned curve for the $i$-th subject by using the warping function $g_{i}^{-1}(t)$.

The models defined in Equations (\ref{eq:regist}), (\ref{eq:response}) and (\ref{eq:logist}) are referred to as joint curve registration and classification (JCRC) models.

\subsection{Estimation procedures}
\label{sec:2.2}
In this subsection, we present the estimation procedures for the proposed JCRC models. We start by considering the estimation procedure in the first level model.

Suppose that $n_{ki}$ observations are obtained for the $i$-th curve in the $k$-th group such that $\ve{x}_{aki}=\big(x_{aki}(t_{i1}),\ldots,x_{aki}(t_{in_{ki}})\big)^{\T}$. Then model (\ref{eq:regist}) can be expressed as
\begin{equation}\label{eq:discrete_regist}
\ve{x}_{aki} = \ve{\tau}_{ak}(g_{ki}) + \ve{r}_{aki} + \ve{\epsilon}_{ai}, \hspace{2mm} a = 1, 2; \hspace{2mm} k = 0, 1; \hspace{2mm} i = 1,\dots,N_{k},
\end{equation}
where $\ve{\tau}_{ak}(g_{ki}) = \Big(\tau_{ak}\big(g_{ki}(t_{i1})\big), \dots, \tau_{ak}\big(g_{ki}(t_{in_{ki}})\big)\Big)^{\T}$, $\ve{r}_{aki}$ and $\ve{\epsilon}_{ai}$  are both $n_{ki}$-dimensional column vectors. Let $\ve{S}_{aki}$ and $\ve{H}_{ki}$ be the covariance matrix of $\ve{r}_{aki}$ and $\ve{w}_{ki}$, respectively. Then we approximate the fixed part $\tau_{ak}(t)$ by using basis functions $\{\psi_{1}(t), \dots, \psi_{q}(t)\}$ with weights $\ve{c}_{a} = (c_{a1}, \dots, c_{aq})^{\T}$  for $\zeta_{a}(t)$ and $\ve{d}_{ak} = (d_{ak1}, \dots, d_{akq})^{\T}$ for $\xi_{ak}(t)$, thus $\ve{\tau}_{ak}(g_{ki}) = \ve{\Psi}_{ki}(\ve{c}_{a} + \ve{d}_{ak})$, where $\ve{\Psi}_{ki} = [\ve{\Psi}_{ki1}, \dots, \ve{\Psi}_{kiq}]_{n_{ki} \times q}, \ve{\Psi}_{kil} = (\ve{\psi}_{l}(g_{ki}(t_{i1})), \dots, \ve{\psi}_{l}(g_{ki}(t_{in_{ki}})))^{\T}$ for  $l=1,\ldots,q$.

To this end, all the unknown parameters in model (\ref{eq:discrete_regist}) can be denoted as $\ve{\theta}_x=\{(\ve {c}_a, \ve {d}_{ak}, \ve {w}_k, \ve {w}_{ki},\\ \ve{\rho}_{h},\ve{\rho}_{s},\sigma): i=1,\ldots,N_k; k=0,1; a=1,2\}$. Obviously, model (\ref{eq:discrete_regist}) has a considerable number of parameters, and also has effects that interacts. This renders direct simultaneous likelihood estimation intractable. To address this issue, we apply a scheme proposed by \cite{raket2016separating} in which fixed effects and parameters are estimated and random effects are predicted iteratively on three different levels of modelling. For notational simplicity, we denote $\ve{x}_{ak} = (\ve{x}_{ak1}^{\T}, \dots, \ve{x}_{akn_{k}}^{\T})^{\T} \in \mathbb{R}^{n_{k}}$, where $n_{k} = \sum_{i = 1}^{N_{k}}n_{ki}$, and $\ve{x}_{a} = (\ve{x}_{a0}^{\T}, \ve{x}_{a1}^{\T})^{\T} \in \mathbb{R}^{n}$, where $n = \sum_{k=0}^{1}n_{k}$.  Let $\sigma^{2}\ve{S}_{ak}$, $\sigma^{2}\ve{S}_{a}$ be the covariance matrices of $\ve{r}_{ak} = (\ve{r}_{aki})_{i}$ and $\ve{r}_{a} = (\ve{r}_{ak})_{k}$, respectively. In order to simplify the likelihood computations, all the random effects are scaled by a  noise standard deviation $\sigma$.  The norm induced by a full-rank covariance matrix $\ve{B}$ is denoted by $\|\ve{A}\|^{2}_{\ve{B}} = \ve{A}^{T}\ve{B}^{-1}\ve{A}$. Then the estimation procedure can be implemented iteratively through the following three conditional models.

\subsubsection{Nonlinear model}
In this model, we consider the conditional likelihood estimation of the subject-specific warping functions $\ve{w}_k$ and prediction of the random warping functions $\ve{w}_{ki}$. Given the estimations of $\hat{\ve{c}}_{a}$  and $\hat{\ve{d}}_{ak}$ from the last iteration, the joint probability density function of $(\ve{x}_{aki}, \ve{w}_{ki})$ is given by
\[
\begin{array}{ll}
f(\ve{x}_{aki}, \ve{w}_{ki}) &= f(\ve{x}_{aki}|\ve{w}_{ki})f(\ve{w}_{ki})\\
&\sim N_{n_{ki}}(\ve{\Psi}_{ki}(\hat{\ve{c}}_{a} + \hat{\ve{d}}_{ak}), \ve{I}_{n_{ki}}+\ve{S}_{aki}) \times N_{n_{w}}(\ve{0}, \ve{H}_{ki}).\\
\end{array}
\]

Then, we can simultaneously estimate the fixed warping effects $\ve{w}_{k}$ and predict the random warping effects $\ve{w}_{ki}$ from the following joint conditional log-likelihood function
\begin{equation}\label{eq:est_warping}
\ell(\ve{w}_{k}, \ve{w}_{ki}) = \sum_{a=1}^{2}\sum_{i=1}^{N_{k}}\|\ve{x}_{aki} - \ve{\Psi}_{ki}(\hat{\ve{c}}_{a} + \hat{\ve{d}}_{ak})\|^{2}_{\ve{I}_{n_{ki}}+\ve{S}_{aki}} + 2\sum_{i = 1}^{N_{k}}\|\ve{w}_{ki}\|^{2}_{\ve{H}_{ki}}.
\end{equation}

\subsubsection{Fixed warp model} Given the estimation of $\ve{w}_k$ and the predicted value of $\ve{w}_{ki}$, we have $\ve{x}_{aki} \sim N_{n_{ki}}(\ve{\Psi}_{ki}(\ve{c}_{a} + \ve{d}_{ak}), \ve{I}_{n_{ki}}+\ve{S}_{aki})$ for $a = 1, 2; i = 1, \dots, N_{k}$, where $\ve{I}_{n_{ki}}$ denotes the $n_{ki} \times n_{ki}$ identity matrix, $\ve {c}_a$ is the weights for $\zeta_a$ and $\ve {d}_{ak}$ is the weights for $\xi_{ak}$. Since we assume that $\xi_{ak}$ is centered around $\zeta_{a}$, then we have $\sum_i\ve {d}_{aki}=\ve {0}\in \mathbb{R}^q$. Then the log-likelihood function for the weights $\ve{c}_{a}$ is proportional to
\[
\ell(\ve{c}_{a}) = \sum_{k = 0}^{1}\sum_{i = 1}^{N_{k}}\|\ve{x}_{aki} - \ve{\Psi}_{ki}\ve{c}_{a}\|^{2}_{\ve{I}_{n_{ki}}+\ve{S}_{aki}}.
\]
Leading to the estimate
\[
\hat{\ve{c}}_{a} = (\ve{\Psi}^{\T} (\ve{I}_{n} + \ve{S}_{a})^{-1}{\ve {\Psi}})^{-1}{\ve{\Psi}}^{\T} (\ve{I}_{n} + \ve{S}_{a})^{-1}\ve{x}_{a},
\]
where $\ve{\Psi} = (\ve{\Psi}^{\T}_{01},\dots,\ve{\Psi}^{\T}_{0N_{0}}, \ve{\Psi}^{\T}_{11},\dots, \ve{\Psi}^{\T}_{1N_{1}})^{\T} \in \mathbb{R}^{n \times q}$, $\ve{x}_a=(\ve{x}^{\T}_{a1},\ldots,\ve{x}^{\T}_{aN})^{\T}$.

Given the estimation of $\hat{\ve{c}}_{a}$, the penalized log-likelihood function for the weights $\ve{d}_{ak}$ is proportional to
\[
\ell(\ve{d}_{ak}) =  \sum_{i = 1}^{N_{k}}\|\ve{x}_{aki} - \ve{\Psi}_{i}(\hat{\ve{c}}_{a} + \ve{d}_{ak})\|^{2}_{\ve{I}_{n_{ki}}+\ve{S}_{aki}} + \lambda\ve{d}_{ak}^{\T}\ve{d}_{ak},
\]
Then the maximum likelihood estimation $\hat{\ve d}_{ak}$ is given by
\[
\hat{\ve{d}}_{ak} =  (\ve{\Psi}_{k}^{\T} (\ve{I}_{n_{k}} + \ve{S}_{ak})^{-1} \ve{\Psi}_{k} + \lambda \ve{I}_{q})^{-1}\ve{\Psi}_{k}^{\T}(\ve{I}_{n_{k}} + \ve{S}_{ak})^{-1}(\ve{x}_{ak} - \ve{ \Psi}_{k}\hat{\ve{c}}_{a}).
\]

\subsubsection{Linearized model} In this model, we consider the first-order Taylor approximation of model (\ref{eq:discrete_regist})  at a given prediction $\ve{w}_{ki}^{0}$ ($\ve{w}^{0}_{ki}$ is specified by the estimation of $\ve{w}_{ki}$ from the \textit{nonlinear model} in the current iteration), we can then write this model as a vectorized linear mixed-effects model as
\begin{equation}\label{eq:mod3}
\ve{x}_{a} \approx \ve{G}_{a} + \ve{B}_{a}(\ve{W} - \ve{W}^{0}) + \ve{r}_{a} + \ve{\epsilon},\hspace{1cm} a = 1, 2,
\end{equation}
where $\ve{x}_{a} = \{\ve{x}_{ai}, i = 1, \dots, N\}$,  with effects given by
\[
\begin{array}{lll}
\ve{G}_{a} &=& \Big\{\ve{\Psi}_{ki}|_{g_{ki} = g_{ki}^{0}}(\ve{c}_{a} + \ve{d}_{ak})\Big\}_{kij} \in \mathbb{R}^{n},\\
\ve{B}_{a}  &=& \rm diag(\ve{B}_{aki})_{ki} \in \mathbb{R}^{n \times Nn_{w}},\\
\ve{B}_{aki}  &=&   \bigg\{\partial_{g_{ki}}\Big(\tau_{ak}\big(g_{ki}(t_{j})\big)\Big)\Big|_{g_{ki}=g_{ki}^{0}}\Big(\nabla_{\ve{w}_{ki}}\big(g_{ki}(t_{j})\big)\Big)^{\T}\Big|_{\ve{w_{ki}} = \ve{w_{ki}}^{0}}\bigg\}_{j} \in \mathbb{R}^{n_{ki} \times n_{w}},\\
\ve{W}  &=&  (\ve{w}_{ki})_{ki} \sim N_{Nn_{w}}(0, \sigma^{2}\ve{I}_{N}\otimes \ve{H}_{n_{w} \times n_{w}}), \hspace{3mm} \ve{W}^{0}  =  (\ve{w}^{0}_{ki})_{ki} \in \mathbb{R}^{Nn_{w}},\\
\ve{r}_{a} &\sim& N_{n}(0,\sigma^{2}\ve{S}_{a}), \hspace{3mm} \ve{S}_{a} = \rm diag(\ve{S}_{aki})_{ki} \in \mathbb{R}^{n \times n},\\
\ve{\epsilon} &\sim& N_{n}(0,\sigma^{2}\ve{I}_{n}),
\end{array}
\]
where $g^{0}_{ki}(t) = t + w_{k}(t) + w^{0}_{ki}(t)$, $\rm diag(\ve{B}_{aki})_{ki}$ is the block diagonal matrix with the $\ve{B}_{aki}$ matrices along its diagonal, so is $\rm diag(\ve{S}_{aki})_{i}$.  The derivation of the linearized model (\ref{eq:mod3}) is given by \cite{zeng2019}.
The log likelihood function for the model (\ref{eq:mod3}) is proportional to
\[
\ell(\sigma^{2}, \ve{\rho}_{s}, \ve{\rho}_{h}) = \sum_{a = 1}^{2}\sigma^{2}\|\ve{x}_{a} - \ve{G}_{a} + \ve{B}_{a}\ve{W}^{0}\|^{2}_{\ve{V}_{a}} + \sum_{a=1}^{2}\log \rm det\ve{V}_{a} + 2 n\log\sigma^{2},
\]
where $\ve{V}_{a} = \ve{S}_{a} + \ve{B}_{a}(\ve{I}_{n} \otimes \ve{H}_{n_{w} \times n_{w}})\ve{B}_{a}^{\T} + \ve{I}_{n}$.
Then all the variance parameters can be estimated by maximizing the log-likelihood function.

We then consider the estimation of the second level model (\ref{eq:response}) and (\ref{eq:logist}). The implementation of the above estimation procedure for the first level model yields
\begin{equation}\label{eq:logis2}
\rm{logit}(\pi_{i}) = b_{0} + \ve{v}_{i}^{\T}\ve{b_{1}} +\int \ve{x}_{i}(\hat{g}_{i}^{-1}(t))\ve{\beta}(t)dt, \hspace{5mm}i = 1,\dots,N.
\end{equation}

Motivated by the fast fitting methods for generalized functional linear models proposed by \cite{goldsmith2011}, $x_{ai}(\hat{g}_{i}^{-1}(t))$ can be approximated using a finite series expansion as follows
\[
x_{ai}(\hat{g}_{i}^{-1}(t))  = \sum_{l =1}^{K_{x}}p_{ail}\phi_{al}(t)=  \ve{p_{ai}}^{\T}\ve{\phi}_{a}(t),\hspace{5mm} \rm{for}\hspace{5mm}a = 1, 2,
\]
where $p_{ail} = \int x_{ai}(\hat{g}_{i}^{-1}(t))\phi_{al}(t)dt$, $\ve{p}_{ai}=(p_{ai1},\ldots,p_{aiK_{x}})^{\T}$ and  $\ve{\phi}_{a}(t) = \big(\phi_{a1}(t), \dots, \phi_{aK_{x}}(t)\big)^{\T}$ is the collection of  the first $K_{x}$ eigenfunctions of the smoothed covariance matrix $K_{\hat{x}_{a}}(s, t) = \rm{Cov}[\hat{x}_{ai}(s), \hat{x}_{ai}(t)]$ \citep{ram05}. Generally, the functional coefficient $\ve{\beta}_{a}(t)$ can also be approximated in a similar way by using a set of basis functions, for instance, the same eigenfunctions of the smoothed covariance $K_{\hat{x}_{a}}(s, t)$, wavelet basis, Fourier basis and spline basis, etc. In this paper,  we apply a truncated power series spline expansion for $\ve{\beta}_a(t)$ due to it's computational efficiency. The resulting estimation procedure can be easily applied to the case with other basis functions. A truncated power series spline expansion for $\ve{\beta}_a(t)$ is expressed as
\begin{equation}
\ve{\beta}_{a}(t)  = e_{a1} + e_{a2}t + \sum_{l=3}^{K_{e}}e_{al}(t - \kappa_{l})_{+}
= \ve{\varphi}^{\T}_{a}(t)\ve{e}_{a},\hspace{5mm} \rm{for}\hspace{5mm}
a = 1, 2, \label{eq:functional_coef}
\end{equation}
where $K_{e}$ is the number of truncated power series spline basis, $\kappa_{l}$ is the location of the $l$-th knot and without loss of generality is taken to be the quantile of the unique data set ${\cal T}=\{t_{ij}:i=1,\ldots,N; j=1,\ldots,n_i\}$. $\ve{e}_a=(e_{a1},\ldots,e_{aK_{e}})^{\T}$ is a vector of parameters, $\ve{\varphi}_a(t)=\big(1,t,(t-\kappa_{1})_+,\ldots,(t-\kappa_{K_{e}})_+\big)^{\T}$ is the basis spline with $t_{+}=\max(t,0)$. To induce smoothing, we assume that $\{e_{al}\}_{l=3}^{K_{e}}\sim N_{K_{e} - 2}(\ve{0}, \sigma_{e}^2\ve{I})$.

Thus, the integral in model (\ref{eq:logis2}) becomes
\[
\int x_{ai}(\hat{g}_{i}^{-1}(t))\ve{\beta}_{a}(t)dt  = \int\ve{p_{ai}}^{\T}\ve{\phi}_{a}(t)\ve{\varphi}_{a}^{\T}(t)\ve{e}_{a}dt = \ve{p}_{ai}^{\T}\ve{J}_{a\phi\varphi}\ve{e}_{a}, \hspace{5mm} \rm{for} \hspace{5mm} a = 1, 2,
\]
where $\ve{J}_{a\phi\varphi}$ is a $K_{x}\times K_{e}$ dimensional matrix with the $(l,s)$-th entry equal to $\int \phi_{al}(t)\varphi_{as}(t)dt$ \citep{ram05}. Then Equations (\ref{eq:response}) and (\ref{eq:logis2}) can be reformulated as
\begin{equation}\label{eq:glmm}
\begin{array}{lll}
y_{i}|\ve{v}_{i}, \ve{x}_{i}(t)  \sim  \rm{Bernoulli}(\pi_{i}), \\
\rm logit(\pi_{i}) =  b_{0} + \ve{v}_{i}^{\T}\ve{b_{1}} + \sum_{a = 1}^{2}\sum_{l=1}^{2}\ve{p}^{\T}_{al}[J_{a\phi\varphi}]_{\cdot l}e_{al} + \sum_{a = 1}^{2}\sum_{l=3}^{K_{e}}\ve{p}^{\T}_{al}[J_{a\phi\varphi}]_{\cdot l}e_{al},
\end{array}
\end{equation}
where $[J_{a\phi\varphi}]_{\cdot l}$ is the $l$-th column vector of the matrix $[J_{a\phi\varphi}]_{K_{x}\times K_{e}}$. Obviously, the second level model is expressed as a generalized linear mixed effects model (GLMM) with random effects $\{e_{al}\}_{l=3}^{K_{e}}\sim N_{K_{e} - 2}(\ve{0}, \sigma_{e}^2\ve{I})$. Denote all the unknown parameters in this model as $\ve{\theta}_y=\{b_0,\ve{b}_1,e_{11},e_{12},e_{21},e_{22},\sigma_e\}$, then the estimation of $\ve{\theta}_y$ can be obtained by using standard mixed effects software \citep{ruppert2002,mcculloch2008}.

So far, we have discussed the estimation procedure for the JCRC models. Note that, the tuning parameters $K_x$ and $K_e$ should be appropriately selected in order to obtain a satisfactory estimation of the JCRC models. Choices of $K_x$ and $K_e$ have been extensively studied in functional and smoothing literatures, respectively. Following \cite{ruppert2002}, we choose the number of knots $K_e$ large enough to prevent undersmoothing and choose $K_x$ large enough to satisfy the identifiability constraint $K_x\geq K_e$. $K_x$ and $K_e$ can be selected by Cross-Validation method.

\subsection{Prediction}
\label{sec:2.3}
It is of practical interest to predict $y^{\ast}$ at a new set of inputs $(\ve{x}^{\ast}(t), \ve{v}^{\ast})$. Since we don't know which class the new $y^{\ast}$ will belong to in advance, and we don't know which type of warping function should be used in Equations (\ref{eq:regist}) and (\ref{eq:warp}). We propose to use the following iterative method.
\begin{enumerate}
	\item[(1)] Initialize $y^{\ast}$ for the iteration.
	By fitting the model (\ref{eq:logist}) without using functional variables, we have
	\[
	\rm{logit}(\pi) = b_{0} + \ve{v}^{\T}\ve{b_{1}}.
	\]
	We initially predict $\pi^{*}$  as $\frac{\rm{exp}\{\hat{b}_{0} + \ve{v}^{*\T}\hat{\ve{b}}_{1}\}}{1 + \rm{exp}\{\hat{b}_{0} + \ve{v}^{*\T}\hat{\ve{b}}_{1}\}}$, where $\{\hat{b}_{0}, \hat{\ve{b}}_{1}\}$ are the  estimators of $\{b_{0}, \ve{b}_{1}\}$ obtained from the observed training data. Then set $y^{*(0)} = 1$ if $\pi^{*} \geq 0.5$ and $y^{*(0)} = 0$ otherwise.
	
	\item[(2)] Calculate $\ve{x}^{\ast}(g^{-1}(t))$ given $y^{\ast(i_{0})}$, where $i_{0}$ indicates the $i_{0}$-th iteration of the algorithm.
	Given the observed curve $\ve{x}^{\ast}(t) = (x_{1}^{\ast}(t), x_{2}^{\ast}(t))^{\T}$ and the estimators $\hat{\ve{\theta}}_y$, the estimate of the subject-specific warping part $\ve{w}_{k\ast}$ can be obtained by minimizing the joint conditional negative log likelihood
	\[l(\hat{\ve{w}}_{k}, \ve{w}_{k\ast}) = \sum_{a=1}^{2}||\ve{x}^{\ast}_{a} - \ve{\Psi}_{k\ast}(\hat{\ve{c}}_{a} + \hat{\ve{d}}_{ak})||^{2}_{\ve{I}_{n_{k\ast}}+\hat{\ve{S}}_{ak\ast}} + 2||\ve{w}_{k\ast}||^{2}_{\hat{\ve{H}}_{ak\ast}}, ~ k = y^{\ast(i_{0})},
	\]
	where $g_{k\ast}(t) = t + \hat{w}_{k}(t) + w_{k\ast}(t)$ and $\ve{\Psi}_{k\ast}$ is determined by $n_{ki}$ discrete values of the inverse of warping function $g_{k\ast}(t)$. $\ve{x}^{\ast}(g^{-1}(t))$ can then be predicted as $\ve{x}^{\ast}(\hat{g}_{k\ast}^{-1}(t))$ where $\hat{g}_{k\ast}(t) = t + \hat{w}_{k}(t) + \hat{w}_{k\ast}(t)$.
	
	\item[(3)] Update $y^{\ast}$. $y^{\ast}$ can be updated as $y^{\ast(i_{0}+1)}$ from the functional logistic regression model (\ref{eq:logist}) given the data $(\ve{x}^{\ast}(\hat{g}_{k\ast}^{-1}(t)), \ve{v}^{\ast})$, where $k = y^{\ast(i_{0})}$.
	
	\item[(4)] Repeat step (2) and step (3) until the value of $y^{\ast}$ and other variables converge.
\end{enumerate}

\subsection{Theoretical properties}
\label{sec:2.4}
In this subsection, we establish the identifiability of model (\ref{eq:regist}) and asymptotic properties of the proposed estimation procedures. For notational simplicity,  absorbing the slope into the scalar variable yields $\ve{v}_i=(1,v_1,\ldots,v_p)^{\T}\in \mathbb{R}^{p+1}$. Suppose that $\{\ve{x}_i(t): t\in\cal{T}\}$ is a zero mean, second order stochastic process with sample paths in the Hilbert space $L^2(\cal{T})$ consists of all square integrable functions on $\cal{T}$, which without loss of generality we take to be $[0,1]$ in this paper. To establish the asymptotic properties of the proposed estimators, we consider a more general case, where given the variables $(\ve{v}_i, \ve{x}_i(t))$, the response variable $y_i$ follows an exponential family with probability density function
\begin{equation}\label{eq:GPFLM}
p(y_i|\ve{v}_i, \ve{x}_i(t))=\exp\big\{y_i\ve{\vartheta}(\ve{v}_i, \ve{x}_i)-{\cal B}[\ve{\vartheta}(\ve{v}_i, \ve{x}_i)]+{\cal C}(y_i)\big\}
\end{equation}
for known functions $\cal{B}$ and $\cal{C}$, where $\ve{\vartheta}$ corresponds to the canonical parameter in parametric generalized linear model. The systematic component of the model is denoted as
\begin{equation}\label{eq:systematic}
h[\mu(\ve{v}_i, \ve{x}_i)]\overset{\triangle}{=}\eta_{i}=\ve{v}^{\T}_i\ve{b}+\int_0^1\ve{x}_i(g^{-1}_i(t))\ve{\beta}(t)dt,
\end{equation}
where $h$ is a known link function, $\ve{b}=(b_0, \ve{b}^{\T}_1)^{\T}$ is a $(p+1)$-dimensional unknown parameter vector and
\[
\mu(\ve{v}_i, \ve{x}_i)=\text{E}(y_i|\ve {v}_i, \ve{x}_i)={\cal B}'[\ve{\vartheta}(\ve {v}_i, \ve{x}_i)].
\]

Obviously, the functional logistic regression model we defined in Section~\ref{sec:2.1} is a special case of the models defined in (\ref{eq:GPFLM}) and (\ref{eq:systematic}) with $\eta_i=\ve\vartheta(\ve {v}_i, \ve {x}_i)=\text{logit}(\pi_i)=-\log(1-\pi_i)$ and ${\cal C}(y_i)=0$.

\begin{remark}\label{remark1}
	The purpose of this subsection is to develop asymptotic theories for the proposed models rather than provide an optimal estimation procedure, so we consider using the same bases to expand $\ve {x}_{ai}$ and $\ve{\beta}_{a}$ for simplicity. In practical applications, $\ve {x}_{ai}$ and $\ve{\beta}_a$ can be approximated by using two different bases (see Equation (\ref{eq:functional_coef}) for instance), and the theoretical results can be established similarly.
\end{remark}

Denote the true values of $\ve{b}$, $\ve{\beta}_a(t)$ and $\ve{\theta}_x$ as $\ve{b}^*$, $\ve{\beta}^*_{a}(t)$ and $\ve{\theta}^*_x$, respectively. Then under assumptions given in the supplementary material, the following theorems hold. Technical proofs and lemmas are given in the supplementary material. For notational simplicity, we omit the subscripts $a$ and $k$ in model (\ref{eq:regist}), then we have
\begin{theorem}[Identifiability]\label{thm:identifiability}
	Let $\{\tau_1(t), \tau_2(t)\}$ be a random elements in $\cal T$ and assume that $\{g_{1i}(t), g_{2i}(t)\}$ are strictly increasing homeomorphisms with probability one, and such that $E(g_{mi})=Id$ for $m=1,2$, where $Id$ is the identity map, i.e. $Id(x)=x$. Then, model (\ref{eq:regist}) is identifiable.
\end{theorem}

In general, model (\ref{eq:regist}) is not identifiable. However, we show in Theorem \ref{thm:identifiability} that under some additional conditions on the warping functions, identifiability of model  (\ref{eq:regist}) can be restored.

\begin{theorem}\label{thm:scalar_asy}
	Under assumptions (S1)-(S6) given in the supplementary material, we have
	$$\sqrt{N}(\hat{\ve {b}}-\ve {b}^*)\rightarrow N(\ve {0}, \ve {\Omega}^{-1}_1\ve{\Omega}_2\ve{\Omega}^{-1}_1)~\text{as}~ N\rightarrow \infty,$$
	where $\ve{\Omega}_1$ and $\ve{\Omega}_2$ are defined in the supplementary material.
\end{theorem}

This theorem shows that the asymptotic normality of the estimated regression coefficients for the scalar variable is achieved whether the functional variable $\ve {x}_i$'s are modelled to the response variable parametrically or nonparametrically.

\begin{theorem}\label{thm:opt_rate}
	Suppose assumptions (S1)-(S6) in the supplementary material hold, then for any $a\in [1,2]$, it follows that
	$$\|\hat{\ve{\beta}}_{a}(t)-\ve{\beta}^*_{a}(t)\|^2=\int_0^1(\hat{\ve{\beta}}_{a}(t)-\ve{\beta}^*_{a}(t))^2dt=O_p(N^{-(2\gamma-1)/(\alpha+2\gamma)}),$$
	where $\alpha$ and $\gamma$ are defined in the supplementary material.
\end{theorem}

This theorem indicates that the existence of the aligned curves in the proposed model doesn't change the rate of convergence of the functional coefficients estimation.

\begin{corollary}\label{thm:cons_pi}
	Suppose that the assumptions given in Theorem~\ref{thm:scalar_asy} and Theorem~\ref{thm:opt_rate} hold. Assume the inverse of the link function $h^{-1}(\cdot)$ in (\ref{eq:systematic}) exists and is thrice continuously differentiable in $\eta_i$. Denote $\pi^*_i$ as the true value of $\pi_i$. Thus, $\hat{\pi}_i-\pi^*_i=O_p(N^{-1/2})$ for all $i=1,\ldots,N$, i.e. $\hat{\pi}_i$ is a consistent estimator of the probability $\pi^*_i$.
\end{corollary}

\begin{theorem}\label{thm:asy_xa}
	Under assumptions (S7)-(S9) given in the supplementary material, we have
	$$N^{1/2}(\hat{\ve{\theta}}_x-\ve{\theta}^*_x)\rightarrow \text{N}(\ve {0}, \ve {A}(\ve{\theta}^*_x)^{-1}B(\ve{\theta}^*_x)A(\ve{\theta}^*_x)^{-1}),$$
	where $A(\ve{\theta}^*_x)$ and $B(\ve{\theta}^*_x)$ are defined in the supplementary material.
\end{theorem}

Theorem~\ref{thm:asy_xa} shows the asymptotic normality of the parameter estimator in the first-level model.

\section{Simulation studies}
\label{sec:3}
In this section, simulation studies are conducted to investigate the finite sample performance of the proposed models.

\subsection{Simulation study 1}
\label{sec:simu1}
In the first simulation study, 100 data sets were generated with sample size of $N=80$, $N=120$ and $N=180$, respectively, through the following three steps.

We First generate $\ve\tau_{aki}(t)$ from
\[
\ve{\tau}_{aki}(t) = \ve{\mu}_{k}(t) + \ve{r}^{0}_{aki}(t) , \hspace{5mm} k = 0, 1; i = 1, \dots, N_k,
\]
with means
\[
\begin{split}
\ve{\mu}_{0}(t) & = \big(\mu_{10}(t), \mu_{20}(t)\big) = \big(0.6\varphi_N(t; 0,1)+0.4\varphi_B(t; 2,3), \sin(2\pi t+0.5)\big), \\
\ve{\mu}_{1}(t) & =  \big(\mu_{11}(t), \mu_{21}(t)\big) = \big(0.5\varphi_N(t; 0.5,0.5)+0.5\varphi_B(t; 3,4), \sin(2\pi t^{1.2}+0.5)\big),
\end{split}
\]
and $\ve{r}^{0}_{aki} = \ve{P}_{0}^{\T}\ve{\Gamma}_{i0}$, where $\varphi_N$ and $\varphi_B$ represent the normal density function and the Beta density function, respectively. $\ve{P}_{0}^{\T}\ve{P}_{0} = \ve{M}_{0}$ where $\ve{M}_{0}$ was created by Mat$\acute{e}$rn covariance function with $\ve{\rho}_{r} = (100, 0.3, 3)$. $\ve{\Gamma}_{i0}$ is a $n_{ki}\times 1$ vector simulated from $N(0, \sigma^{2}_{r})$ with $\sigma_{r}=0.02$. Suppose that the observations for each curve were sampled at equally spaced points in [0,1]. Moreover, in order to investigate the finite sample performance of the estimators under different sampling frequencies $(n_{ki})$, we set $n_{ki}=100$, $n_{ki}=200$ and $n_{ki}=400$, respectively.

Then we generate the curves by adding the warping function. We use B-spline basis functions with 8 knots to model $\ve{\mu}_0(t)$, $\ve{\mu}_{1}(t)$ and $\ve{r}^{0}_{aki}(t)$, which yields
$\ve{\tau}_{aki} = \ve{\Psi}_{ki}(\ve{c}_{a} + \ve{d}_{ak} + \ve{d}_{aki})$,
where $\ve {c}_a$, $\ve {d}_{ak}$ and $\ve{d}_{aki}$ are coefficients and $\sum_l\ve {d}_{al}=0$. For simplicity, we set $g_{ki}(t)=t+\ve {w}_{ki}(t)$ and took Hyman spline with anchor knots $t_w=(0,0.33,0.67,1)$. Assume $\ve{w}_{ki}\sim N_2(\ve {0}, \ve {T}^{\T}_k\ve{\Gamma}_i)$, where $\ve{T}_{k}^{\T}\ve{T}_{k} = \ve{O}_{k}$ with $\ve{O}_{0} = \begin{bmatrix} 10 & 4 \\ 4 & 8\end{bmatrix}$ and $\ve{O}_{1} = \begin{bmatrix} 10 & 8 \\ 8 & 15\end{bmatrix}$, and $\ve{\Gamma}_{i} = (\Gamma_{i1}, \Gamma_{i2})^{\T}$ with $\Gamma_{i1}, \Gamma_{i2}$ being independent $N(0, \sigma^{2}_{w})$ random variables for $i = 1, \dots, N_k$ with $\sigma_{w}=0.005$. Then we have
$$\ve{\tau}_{aki} = \ve{\Psi}^{\ast}_{ki}(\ve{c}_{a} + \ve{d}_{ak} + \ve{d}_{aki}),$$
where $\ve{\Psi}^{\ast}_{ki}$ is the warped basis function. Let $\epsilon_{ai}\sim N(0,\sigma^2)$ with $\sigma=0.02$, then $\ve{x}_{aki}(t)$ can be generated based on model (\ref{eq:regist}).

Finally, the outcomes $y_i$'s can therefore be generated from the following models
\[
\begin{split}
\eta_{i} & = b_{0} + v_{i}b_{1}  + \frac{1}{n_{ki}}\sum_{j =1}^{n_{ki}}\big[\ve x_{1ki}(g^{-1}_{ki}(t_{ij}))\ve\beta_{1}(t_{ij}) +\ve x_{2ki}(g^{-1}_{ki}(t_{ij}))\ve\beta_{2}(t_{ij})\big], k=0,1,\\
\pi_{i} & = \frac{1}{1 + \text{exp}(-\eta_{i})},\\
y_{i} & \sim \text{Bernoulli}(\pi_{i}),~~i = 1,\dots,N,\\
\end{split}
\]
where $b_0=0.1$ and $b_1=-0.5$ are the scalar coefficients while $\ve\beta_1(t)=\cos(2\pi t)$ and $\ve\beta_2(t)=2(t-1)^2$ the functional coefficients. The scalar variable $v_i$ was generated from the following uniform distribution
\[
v_{ki}  \sim \left\{
\begin{array}{ll}
\text{U}(1, 2), \hspace{4mm} i = 1, \dots, N_0, \hspace{2mm}k = 0,\\
\text{U}(0.5, 1.5), \hspace{4mm} i = 1, \dots, N_1, \hspace{2mm}k = 1.
\end{array}
\right.
\]
where $N_0+N_1=N$. Similar to \cite{goldsmith2011}, a rule of thumb $K_x=K_e=35$ was enough in this simulation study.

To evaluate the finite sample performance of the parametric estimator $\hat{\ve {b}}$, we calculated the bias (BIAS) and sample standard deviation (SSD) over 100 replications. For the estimation $\hat{\ve{\beta}}(t)$ of the functional coefficient, we computed the integrated squared bias (ISBIAS) denoted as
$$\text{ISBIAS}(\hat{\ve{\beta}}_a(t))=\int_0^1\bigg[\text{E}(\hat{\ve{\beta}}_a(t))-\ve{\beta}_a(t)\bigg]^2dt$$
and the integrated mean squared error (IMSE)
$$\text{IMSE}(\hat{\ve{\beta}}_a(t))=\int_0^1\text{E}\bigg[\hat{\ve{\beta}}_a(t)-\ve{\beta}_{a}(t)\bigg]^2dt$$
for each $a\in[1,2]$ over 100 replications.  As measures of performance of the predicted warping functions, we also calculated the integrated mean squared error (IMSE) of $\hat{g}(t)$ over 100 replications. These results were demonstrated in Table~\ref{tab:1} and Table~\ref{tab:2}.
\begin{table}[!htbp]
	\centering
	\caption{The BIAS and the SSD of $\hat{b}_0$ and $\hat{\ve {b}}_1$.}
	\label{tab:1}       
	\begin{tabular}{lllllllll}
		\hline\noalign{\smallskip}
		Sample &Sampling&&\multicolumn{2}{c}{BIAS}&&\multicolumn{2}{c}{SSD}\\
		\cline{4-5}\cline{7-8}
		\hspace{5mm}
		size &frequency &&$\hat{b}_{0}$&$\hat{\ve {b}}_{1}$&&$\hat{b}_{0}$&$\hat{\ve {b}}_{1}$\\
		\noalign{\smallskip}\hline\noalign{\smallskip}\\
		&\hspace{5mm}$100$& & 0.67 & 0.54 && 0.89 & 0.69 \\
		$N=80$ & \hspace{5mm}$200$& & 0.75  & 0.60 && 0.96 & 0.77 \\
		& \hspace{5mm}$400$& & 0.61  & 0.51 && 0.79  & 0.63  \\\hline
		& \hspace{5mm}$100$& & 0.53 & 0.43 && 0.68 & 0.54 \\
		$N=120$ & \hspace{5mm}$200$& & 0.56  & 0.45 && 0.69 & 0.53 \\
		&\hspace{5mm}$400$&& 0.56  & 0.44 && 0.72 & 0.57 \\\hline
		& \hspace{5mm}$100$& & 0.48 & 0.38 && 0.59 & 0.47 \\
		$N=180$ & \hspace{5mm}$200$& & 0.40  & 0.32 && 0.52 & 0.41 \\
		& \hspace{5mm}$400$&& 0.48 & 0.35 && 0.60 & 0.44 \\
		\noalign{\smallskip}\hline
	\end{tabular}
\end{table}

\begin{table}[!htbp]
	\centering
	\caption{The ISBIAS and the IMSE of $\hat{\ve{\beta}}_1(t)$, $\hat{\ve{\beta}}_2(t)$ and the IMSE of the predicted warping functions.}
	\label{tab:2}       
	\begin{tabular}{lllllllllll}
		\hline\noalign{\smallskip}
		Sample &Sampling&&\multicolumn{2}{c}{ISBIAS ~($\times 10^{-2}$)}&&\multicolumn{3}{c}{IMSE}\\
		\cline{4-5}\cline{7-10}
		\hspace{5mm}
		size &frequency &&$\hat{\ve {\beta}}_{1}(t)$&$\hat{\ve{\beta}}_{2}(t)$&&$\hat{\ve{\beta}}_{1}(t)$&$\hat{\ve{\beta}}_{2}(t)$&& $\hat{g}(t)~(\times 10^{-3})$\\
		\noalign{\smallskip}\hline\noalign{\smallskip}\\
		&\hspace{5mm}$100$& & 0.51 & 1.03  && 8.13 & 38.44 && 0.18 \\
		$N=80$ & \hspace{5mm}$200$& &  0.21 & 0.42 && 6.65 & 60.48 && 0.18\\
		&\hspace{5mm}$400$&&  0.18  & 0.25  && 8.85 & 20.46 && 0.16\\\hline
		& \hspace{5mm}$100$& &  0.25 & 0.74 && 3.71 & 19.81 && 0.22 \\
		$N=120$ & \hspace{5mm}$200$& &  0.16 & 0.41 && 4.38 & 26.51 && 0.20\\
		&\hspace{5mm}$400$&&  0.07 & 0.02 && 3.57 & 23.40 && 0.17\\\hline
		& \hspace{5mm}$100$& &  0.21 & 0.65 && 3.12 & 13.33 && 0.19\\
		$N=180$ & \hspace{5mm}$200$& &  0.09 & 0.28 && 3.14 & 1.94 && 0.19\\
		&\hspace{5mm}$400$&&  0.04 & 0.12 && 3.02 & 1.22 && 0.16\\
		\noalign{\smallskip}\hline
	\end{tabular}
\end{table}

\begin{figure}[!htbp]
	\centering
	\includegraphics[width=\textwidth,height=60mm]{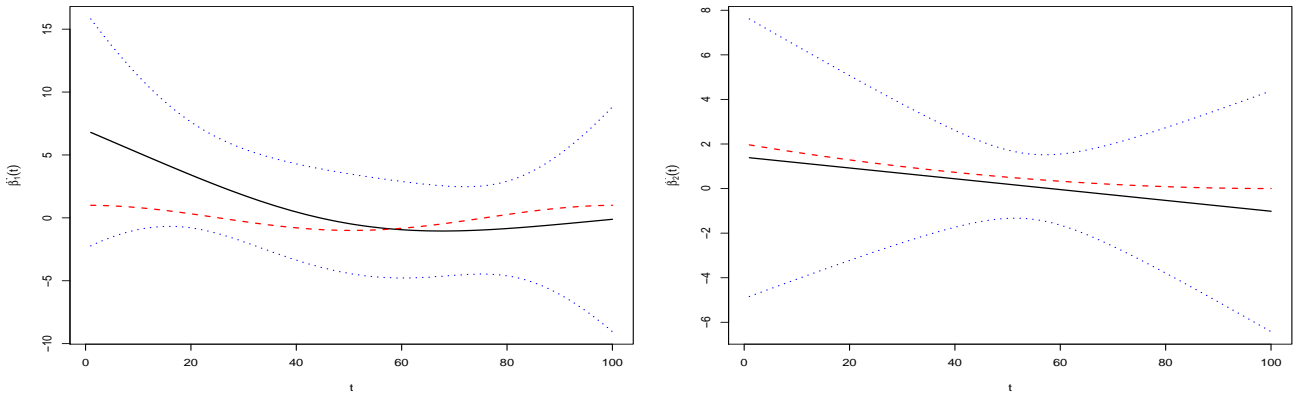}
	\caption{An example of the estimation $\hat{\ve \beta}(t)$ with $N=80$ and $n_{ki}=100$. The solid lines stand for the true $\ve\beta(t)$, the dashed lines stand for the estimation $\hat{\ve\beta}(t)$ and dotted lines represent the 95\% confidence intervals.}
	\label{fig:1}       
\end{figure}

\begin{figure}[!htbp]
	\centering
	\begin{subfigure}[t]{0.4\textwidth}
		\includegraphics[width=\textwidth]{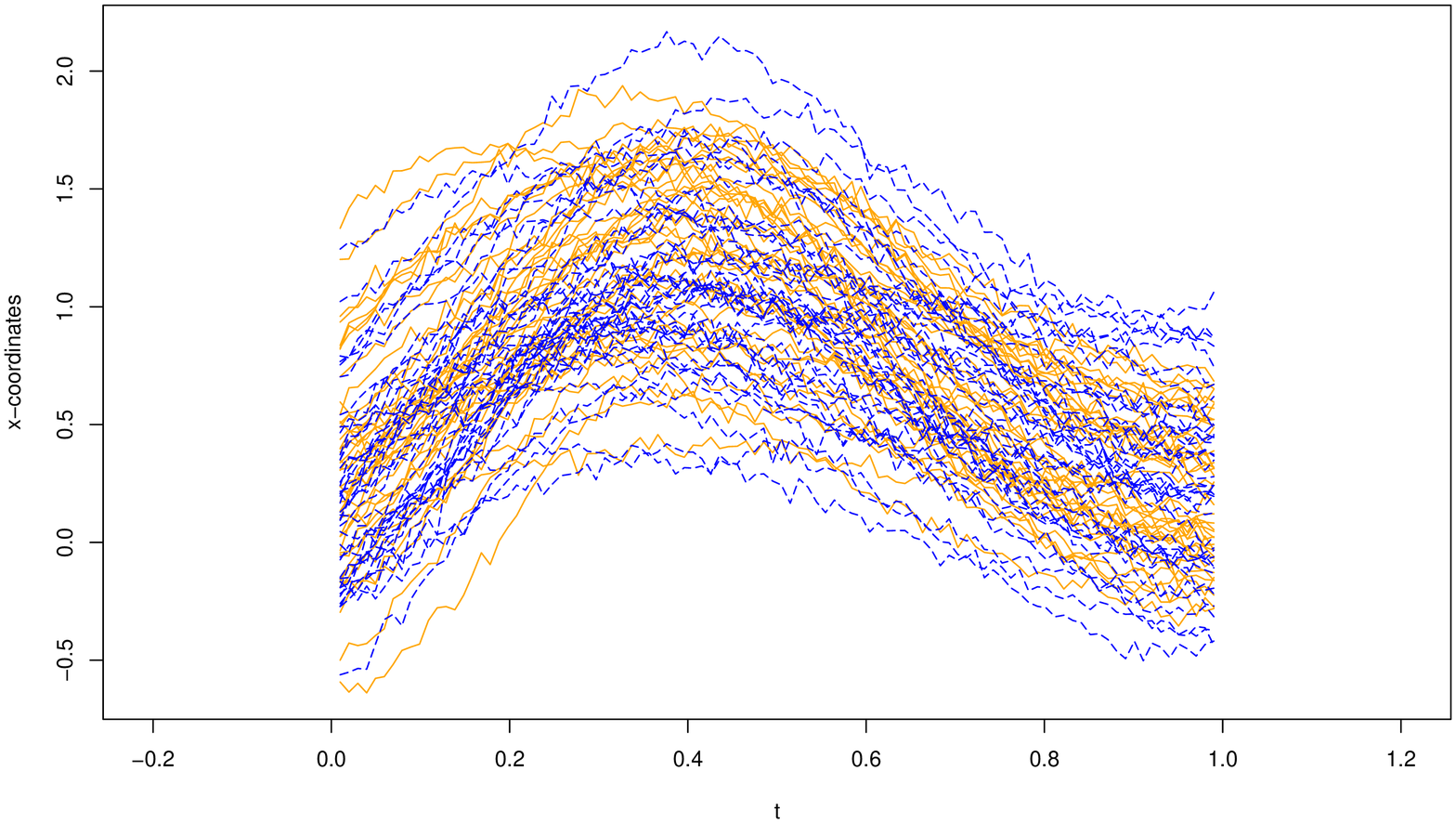}
		\caption{Original $\ve x_1(t)$}
	\end{subfigure}
	\quad
	\begin{subfigure}[t]{0.4\textwidth}
		\includegraphics[width=\textwidth]{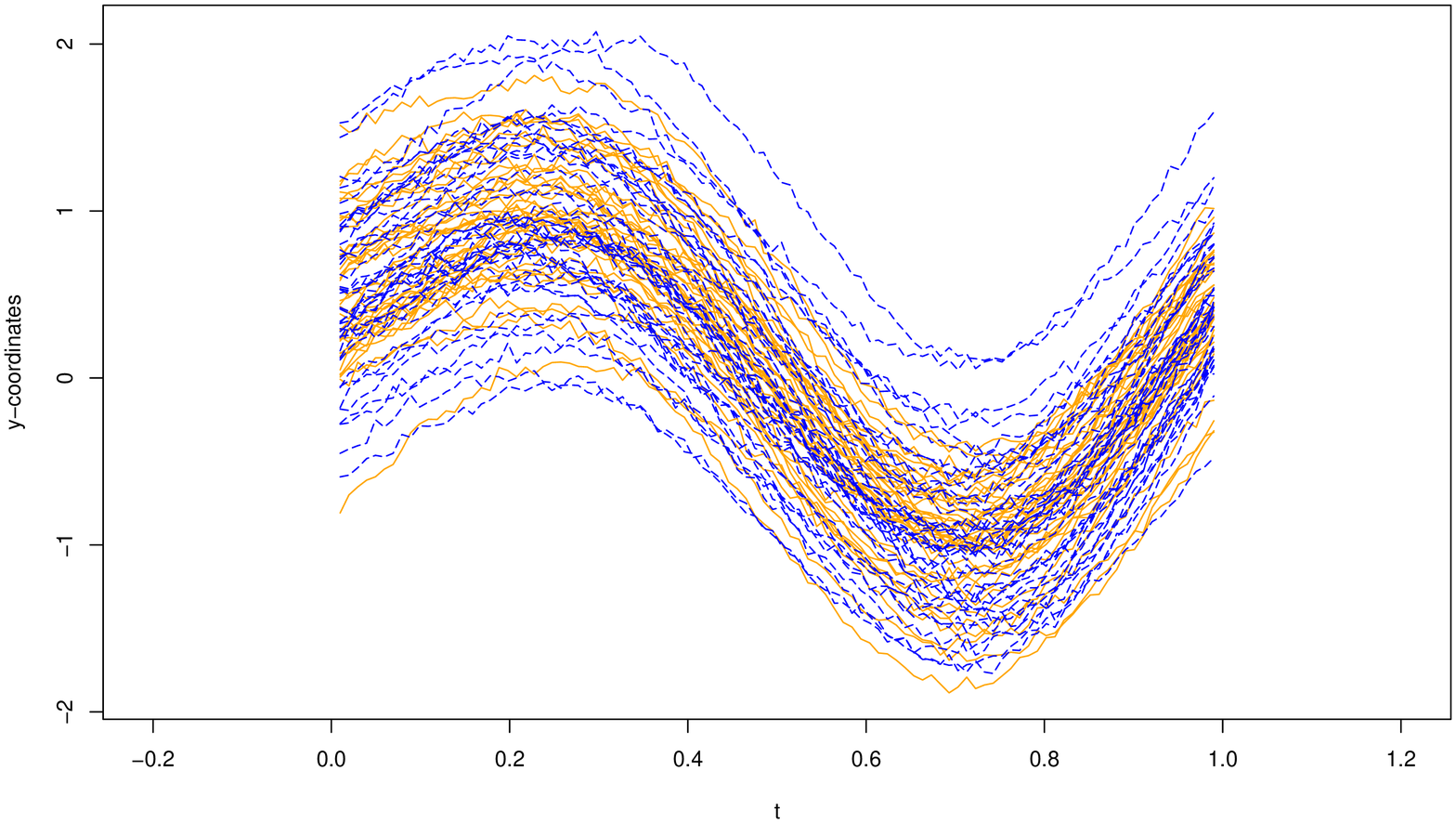}
		\caption{Original $\ve x_2(t)$}
	\end{subfigure}
	\quad
	\begin{subfigure}[t]{0.4\textwidth}
		\includegraphics[width=\textwidth]{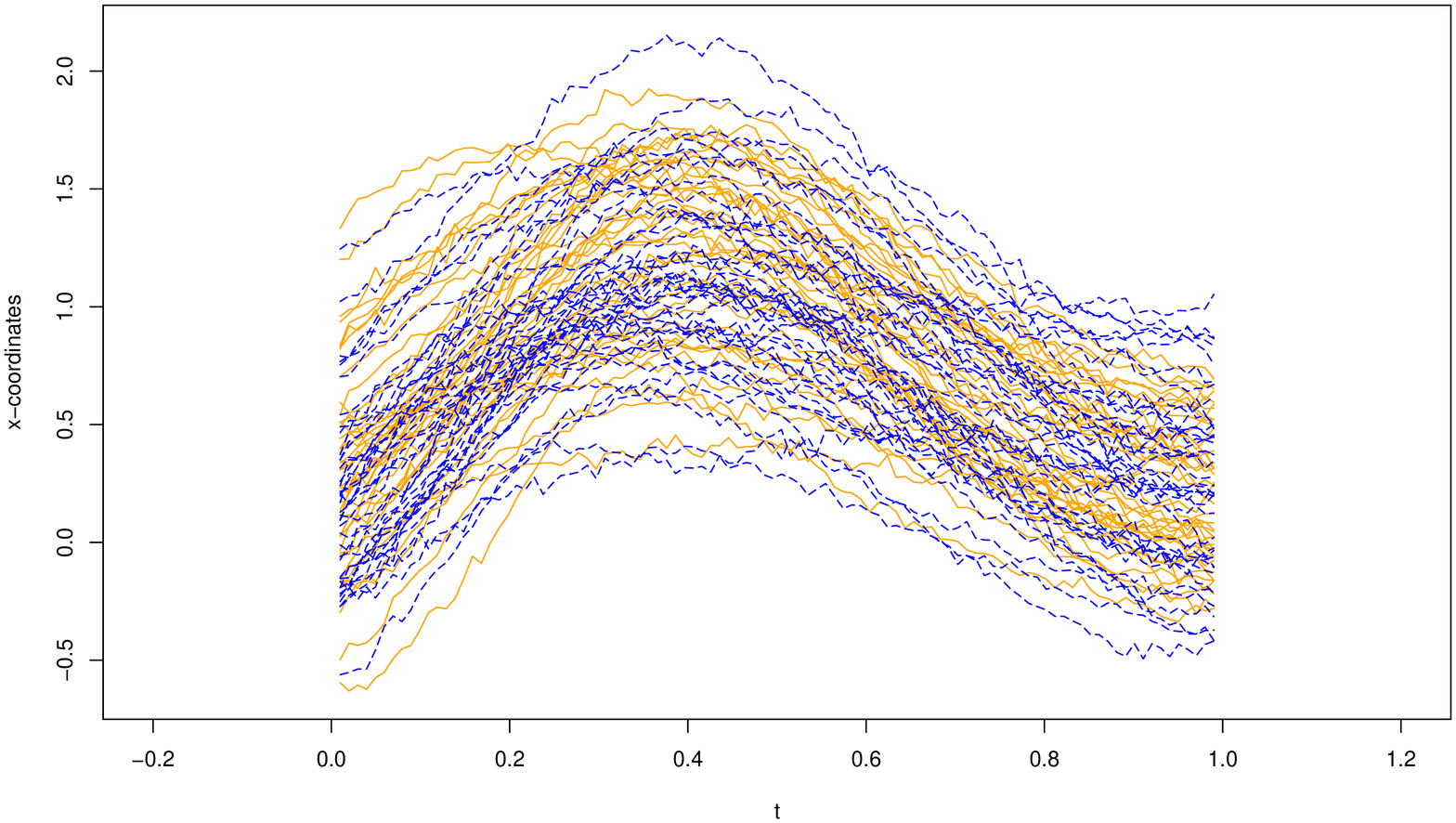}
		\caption{Aligned $\ve x_1(t)$}
	\end{subfigure}
	\quad
	\begin{subfigure}[t]{0.4\textwidth}
		\includegraphics[width=\textwidth]{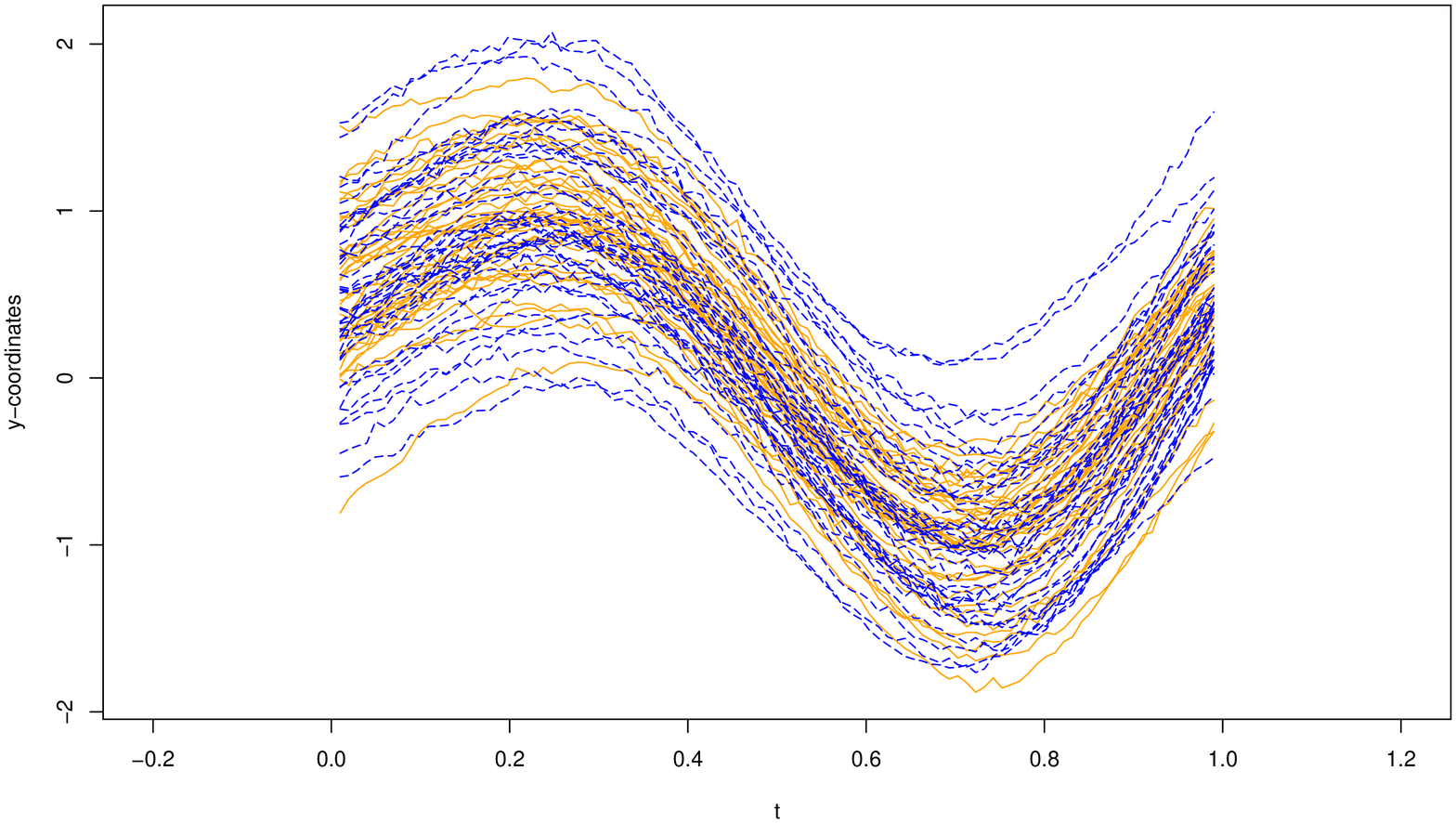}
		\caption{Aligned $\ve x_2(t)$}
	\end{subfigure}
	\caption{An example of raw curves and corresponding aligned curves with $N=80$ and $n_{ki}=100$. Curves in solid orange lines represent the first group $(y=0)$ and the blue dashed lines represent the second group $(y=1)$. }
	\label{fig:2}       
\end{figure}

\begin{figure}[!htbp]
	\centering
	\includegraphics[width=\textwidth,height=60mm]{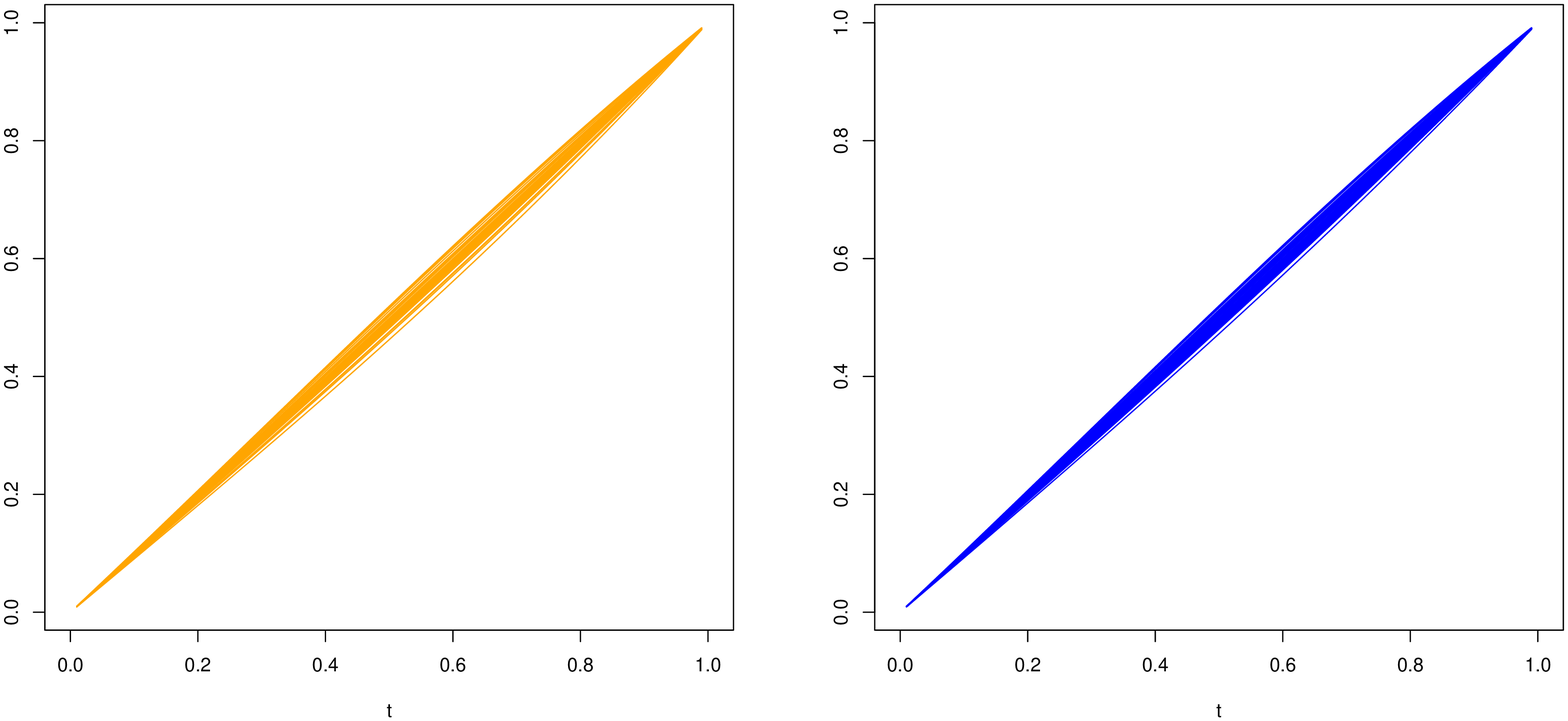}
	\caption{Warping functions corresponding to aligned $\ve x_1(t)$ (left panel) and aligned $\ve x_2(t)$ (right panel).}
	\label{fig:3}       
\end{figure}

Examination of Table~\ref{tab:1} and Table~\ref{tab:2} indicated that (i) the proposed methodology performed quite well in terms of estimation. The values of BIAS for parametric estimators and the values of ISBIAS for the functional coefficient estimators were reasonably small. For a fixed sampling frequency, the BIASs, SSDs, ISBIASs and the IMSEs decreased as the sample size increased. In general, increasing sample size improved the accuracy of estimations as expected; (ii) the sampling frequency didn't has significant influence on the performance of the estimators $\hat{\ve{b}}$ regardless of sample sizes. For $\hat{\ve{\beta}}_1(t)$ and $\hat{\ve{\beta}}_2(t)$, they had better performances at a high sampling frequency for a fixed sample size; (iii) the predicted warping functions performed well in the sense that all the average IMSE values were small regardless of sample sizes and sampling frequencies. Fig.\ref{fig:1} presents an example of the estimation of the functional coefficient $\ve{\beta}(t)$. The registration results were presented in Fig.\ref{fig:2} and Fig.\ref{fig:3} shows the corresponding warping functions to the aligned curves.

\subsection{Simulation study 2}
\label{sec:simu2}
In this subsection, we investigate the performance of the proposed method in terms of prediction. The settings are slightly different to those in simulation study \ref{sec:simu1}. Specifically, the true mean curves were set to be
\[
\begin{split}
\ve{\mu}_{0}(t) & = \big(\mu_{10}(t), \mu_{20}(t)\big) = \big(\text{exp}\{\text{cos}(2\pi t)\}, \text{exp}\{\text{sin}(2\pi t)\}\big), \\
\ve{\mu}_{1}(t) & =  \big(\mu_{11}(t), \mu_{21}(t)\big) = \big(\text{exp}\{\text{cos}(2\pi t^{1.1} - \delta_{1})\}, \text{exp}\{\text{sin}(2\pi t^{1.2} + \delta_{1})\}\big),
\end{split}
\]
and the scalar variables $v_i$'s were sampled from the following uniform distributions
\[
v_{ki}  \sim \left\{
\begin{array}{ll}
\text{U}(1, 2), \hspace{4mm} i = 1, \dots, N_0, \hspace{2mm}k = 0,\\
\text{U}(1 - \delta_{2}, 2 - \delta_{2}), \hspace{4mm} i = 1, \dots, N_1, \hspace{2mm}k = 1,\\
\end{array}
\right.
\]
where $\delta_1$ and $\delta_2$ were introduced to illustrate the degree of overlapping between two groups through functional and scalar variables, respectively.

A total of $N=120$ curves were generated from the above settings, and for each curve,  100 equidistant points $t_j=\frac{j+1}{102}$ ($j=1,\ldots,100$) was used as input grid.  We considered the following two scenarios: (A) $\delta_1 = 0.18$, $\delta_{2} = 0.7$, $4\sigma_{w} = \sigma_{r} = \sigma = 0.03$ ; (B)$\delta_1 = 0.15$, $\delta_{2} = 0.5$, $4\sigma_{w} = \sigma_{r} = \sigma = 0.02$. For each scenario, half of the data was selected as training data and the rest is set to be test data. The performance of classification of our proposed method was evaluated by calculating three criteria, i.e., 	classification accuracy (CA), the Rand index (RI) \citep{rand1971} and  adjusted Rand index (ARI) \citep{hubert1985} for each scenario over 100 replications. We also compared our proposed JCRC method with joint model with only functional variables (denoted by JCRC-f), the logistic linear regression model without functional variables (denoted by LLR), curve classification based on the square-root velocity representation \citep{srivastava2011} (denoted by SRV), the integration of Generalized Procrustes analysis \citep{gower1975} and self-modelling method \citep{gervini2004} (denoted by GPSM).

\begin{figure}[!htbp]
	\centering
	\begin{subfigure}[t]{0.4\textwidth}
		\includegraphics[width=\textwidth]{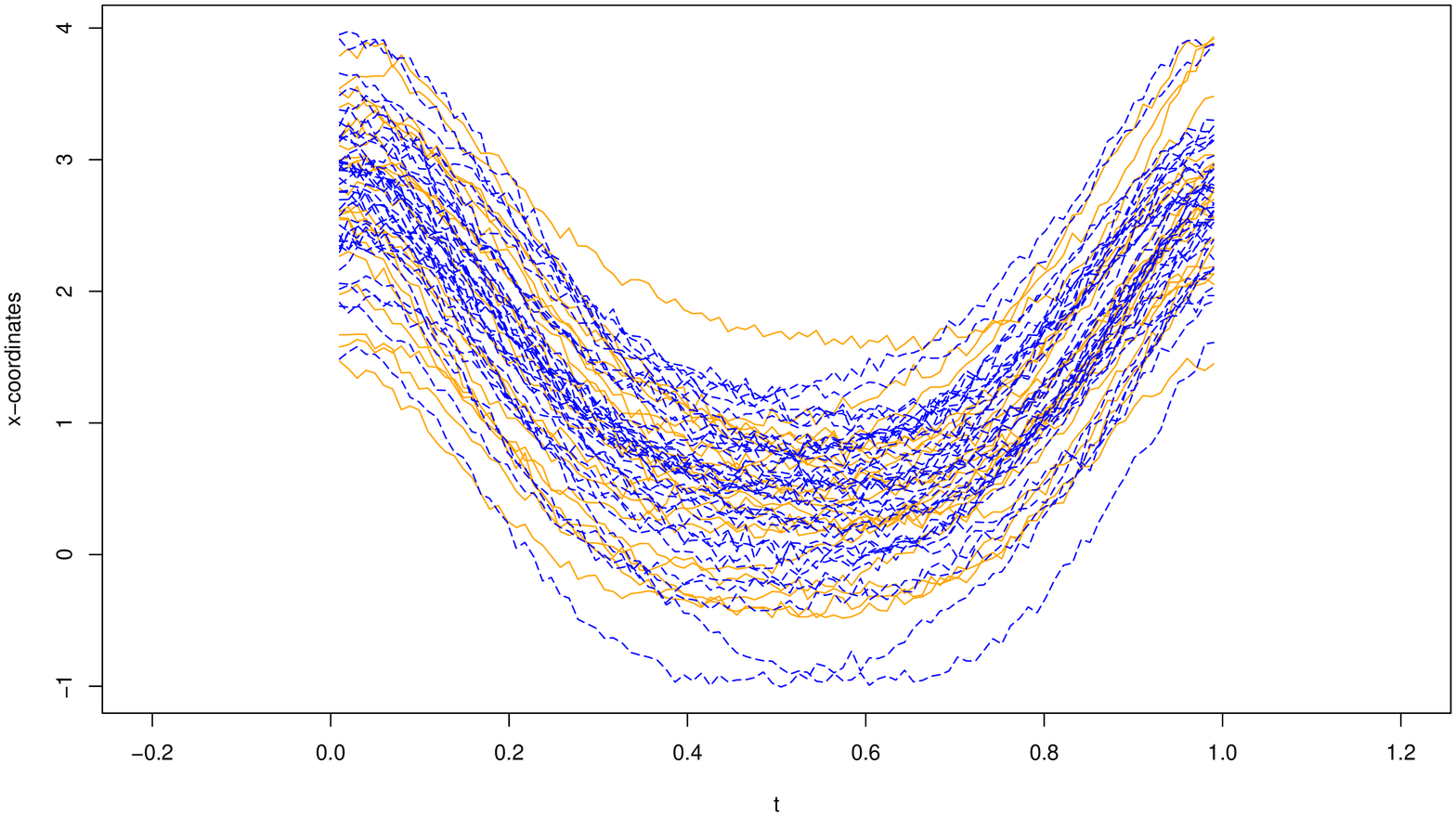}
		\caption{Original $\ve x_1(t)$}
	\end{subfigure}
	\begin{subfigure}[t]{0.4\textwidth}
	\includegraphics[width=\textwidth]{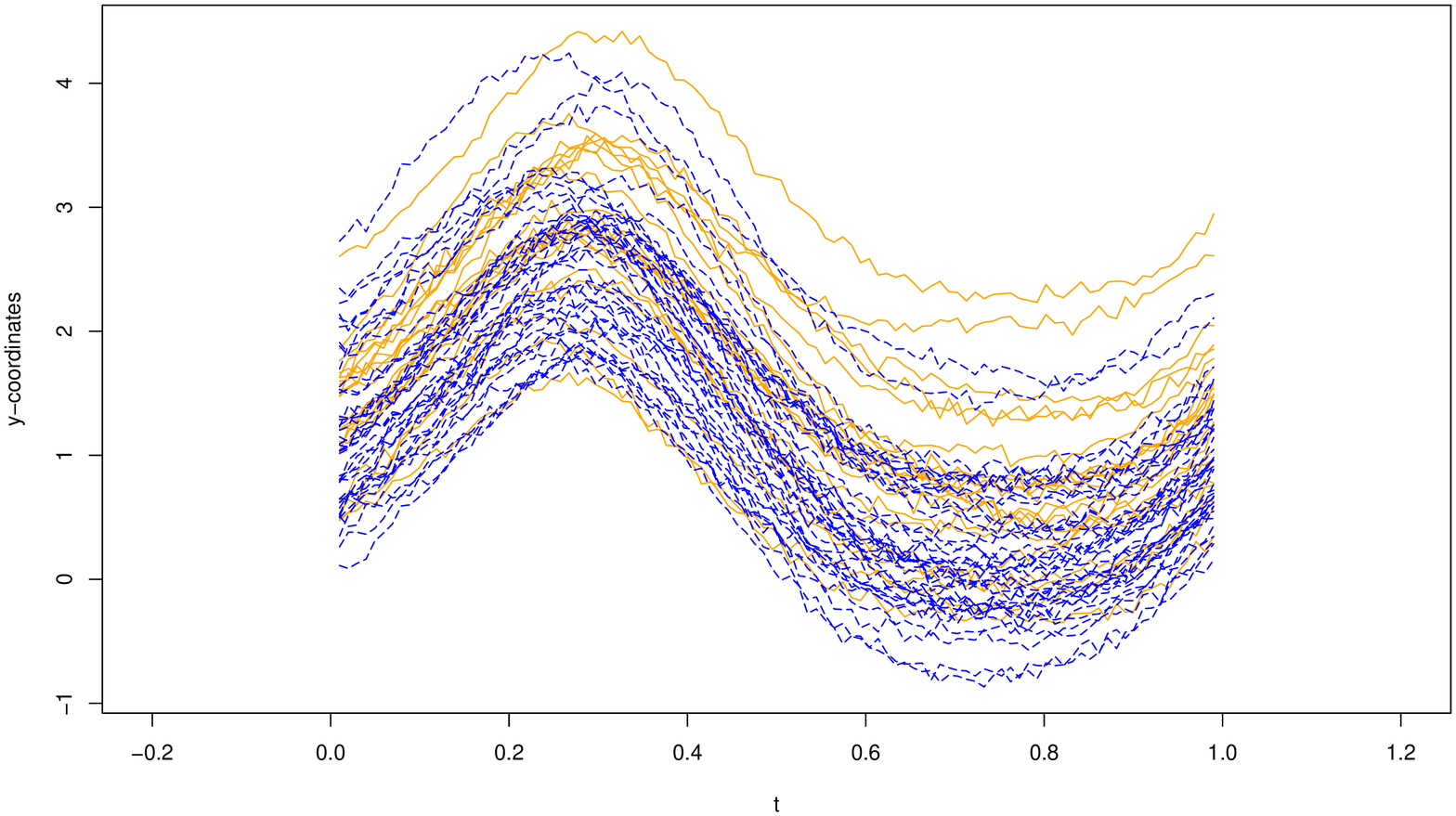}
	\caption{Original $\ve x_2(t)$}
    \end{subfigure}
	\quad
	\begin{subfigure}[t]{0.4\textwidth}
		\includegraphics[width=\textwidth]{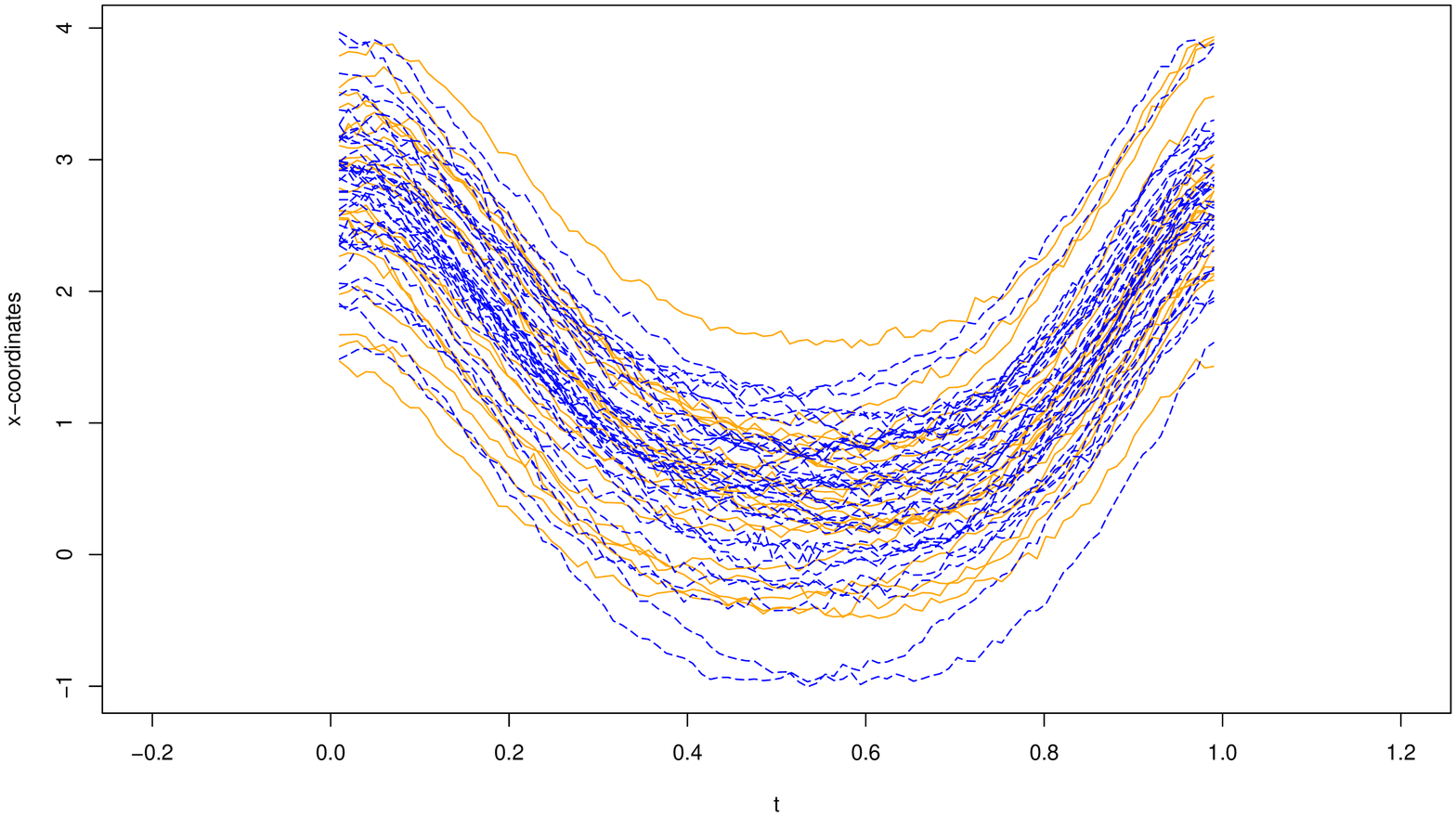}
		\caption{Aligned $\ve x_1(t)$}
	\end{subfigure}
	\quad
	\begin{subfigure}[t]{0.4\textwidth}
		\includegraphics[width=\textwidth]{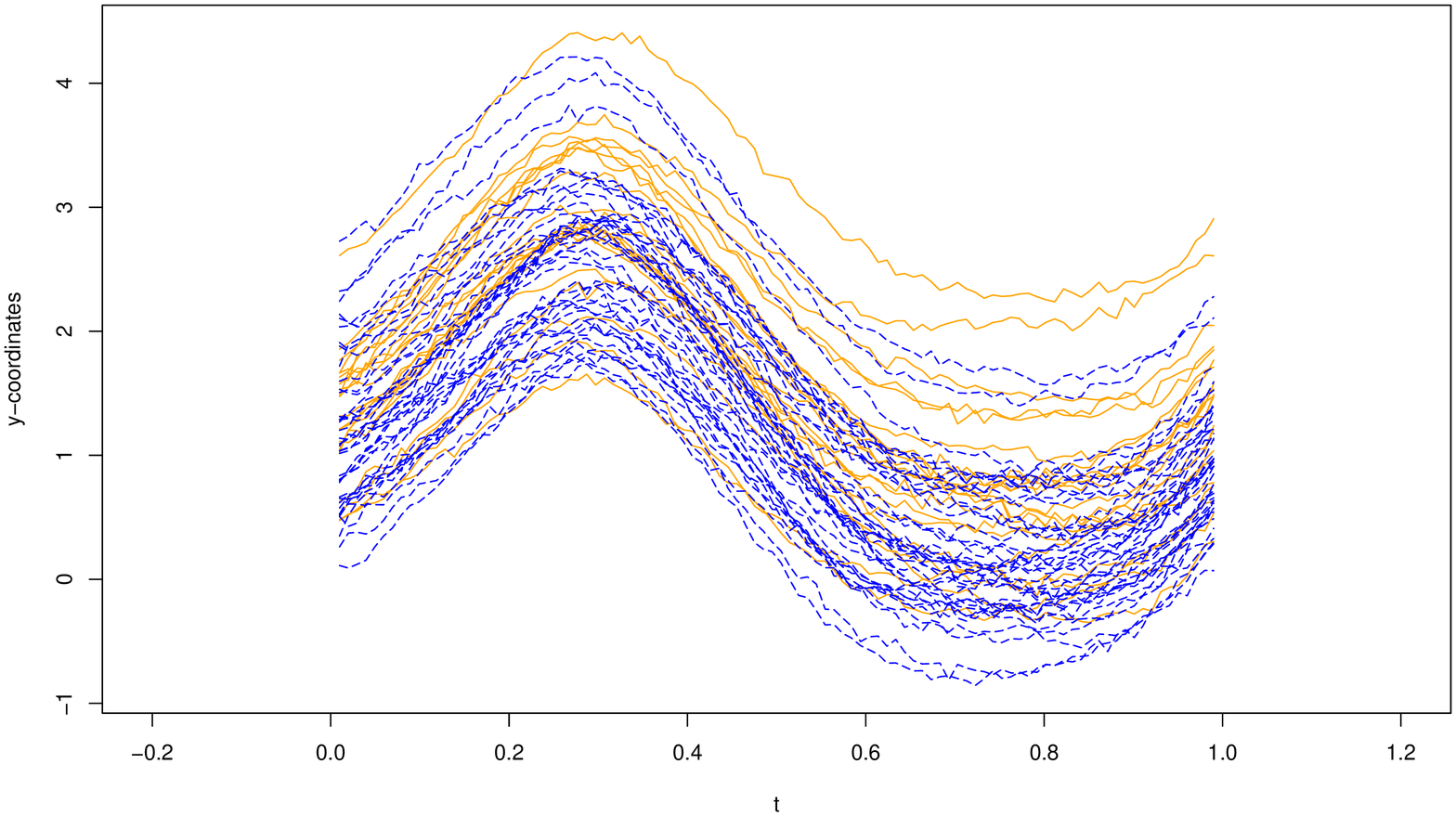}
		\caption{Aligned $\ve x_2(t)$}
	\end{subfigure}
	\caption{An example of raw curves and corresponding aligned curves in Scenario A. Curves in solid orange lines represent the first group $(y=0)$ and the blue dashed lines represent the second group $(y=1)$. }
	\label{fig:4}       
\end{figure}

\begin{figure}[!htbp]
	\centering
	\begin{subfigure}[t]{0.4\textwidth}
		\includegraphics[width=\textwidth]{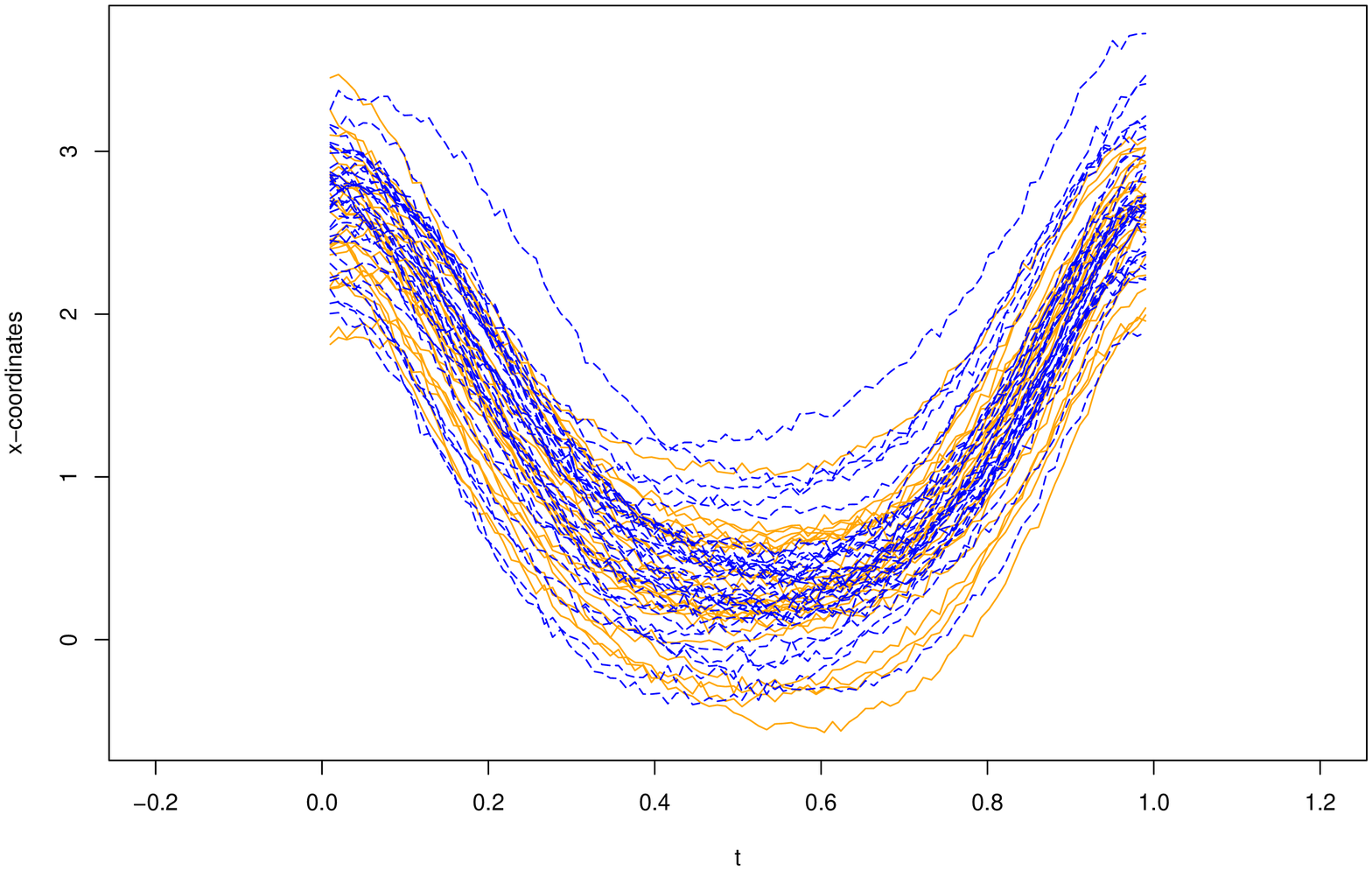}
		\caption{Original $\ve x_1(t)$}
	\end{subfigure}
	\quad
	\begin{subfigure}[t]{0.4\textwidth}
		\includegraphics[width=\textwidth]{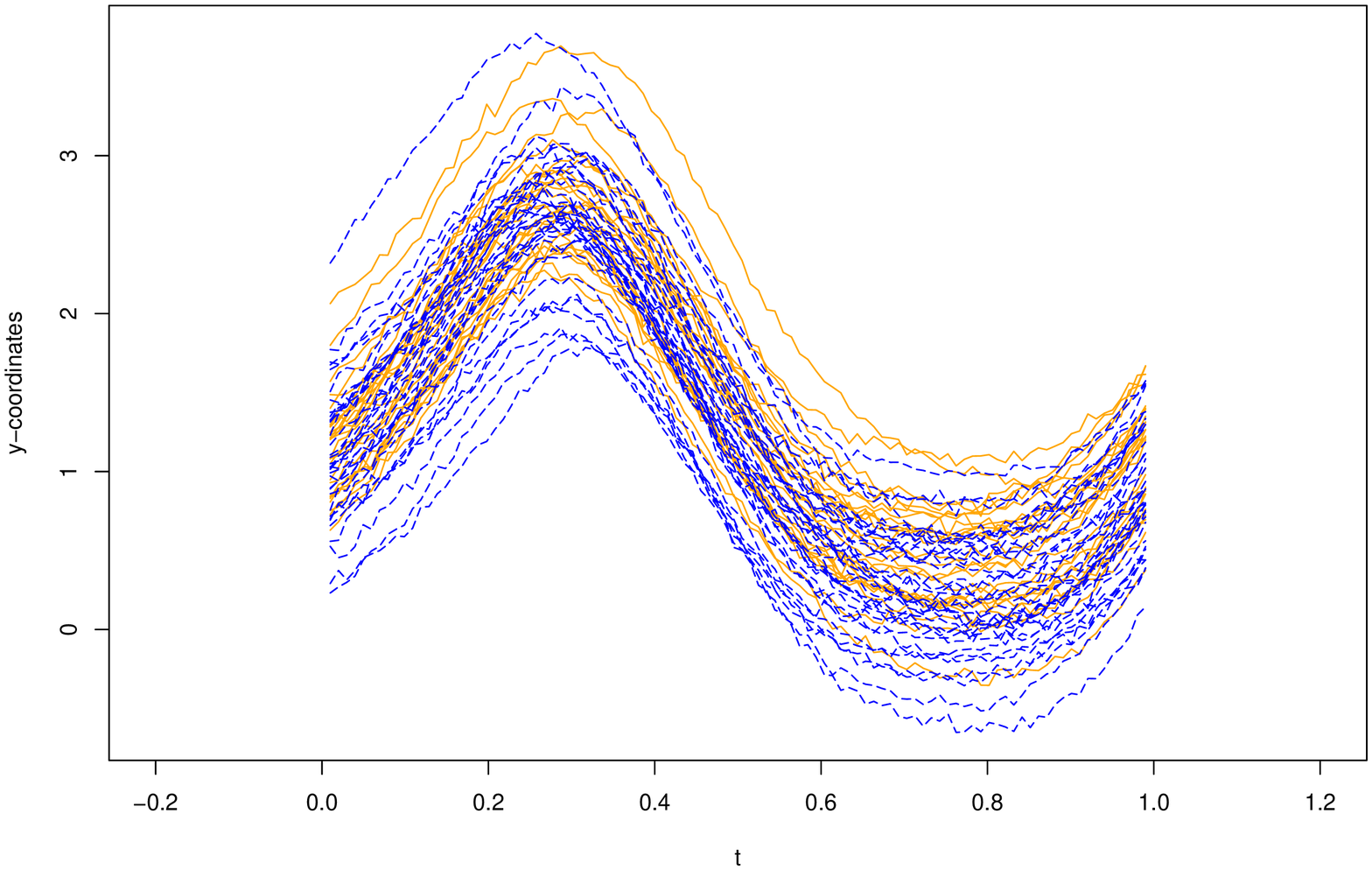}
		\caption{Original $\ve x_2(t)$}
	\end{subfigure}
	\quad
	\begin{subfigure}[t]{0.4\textwidth}
		\includegraphics[width=\textwidth]{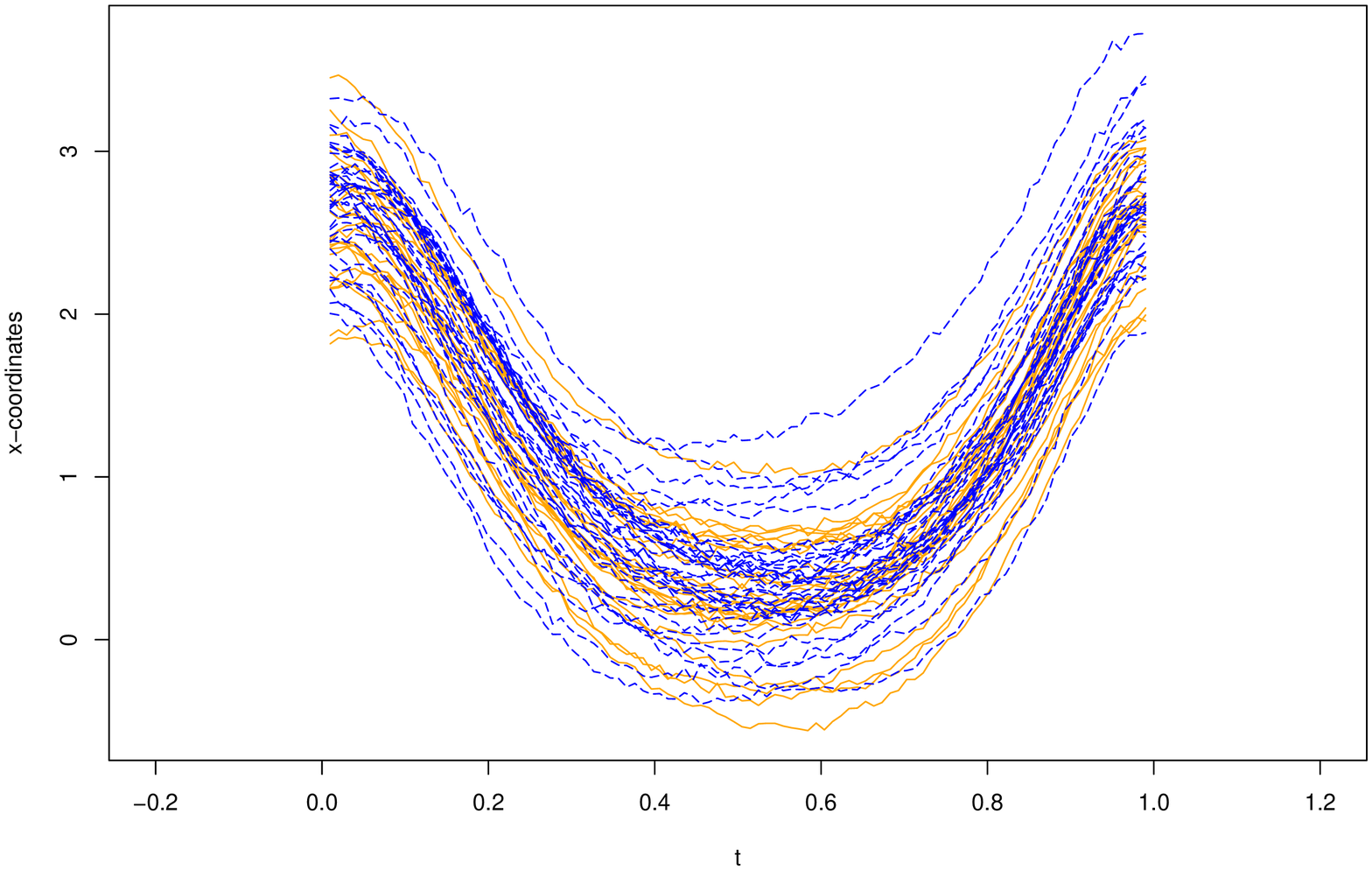}
		\caption{Aligned $\ve x_1(t)$}
	\end{subfigure}
	\begin{subfigure}[t]{0.4\textwidth}
		\includegraphics[width=\textwidth]{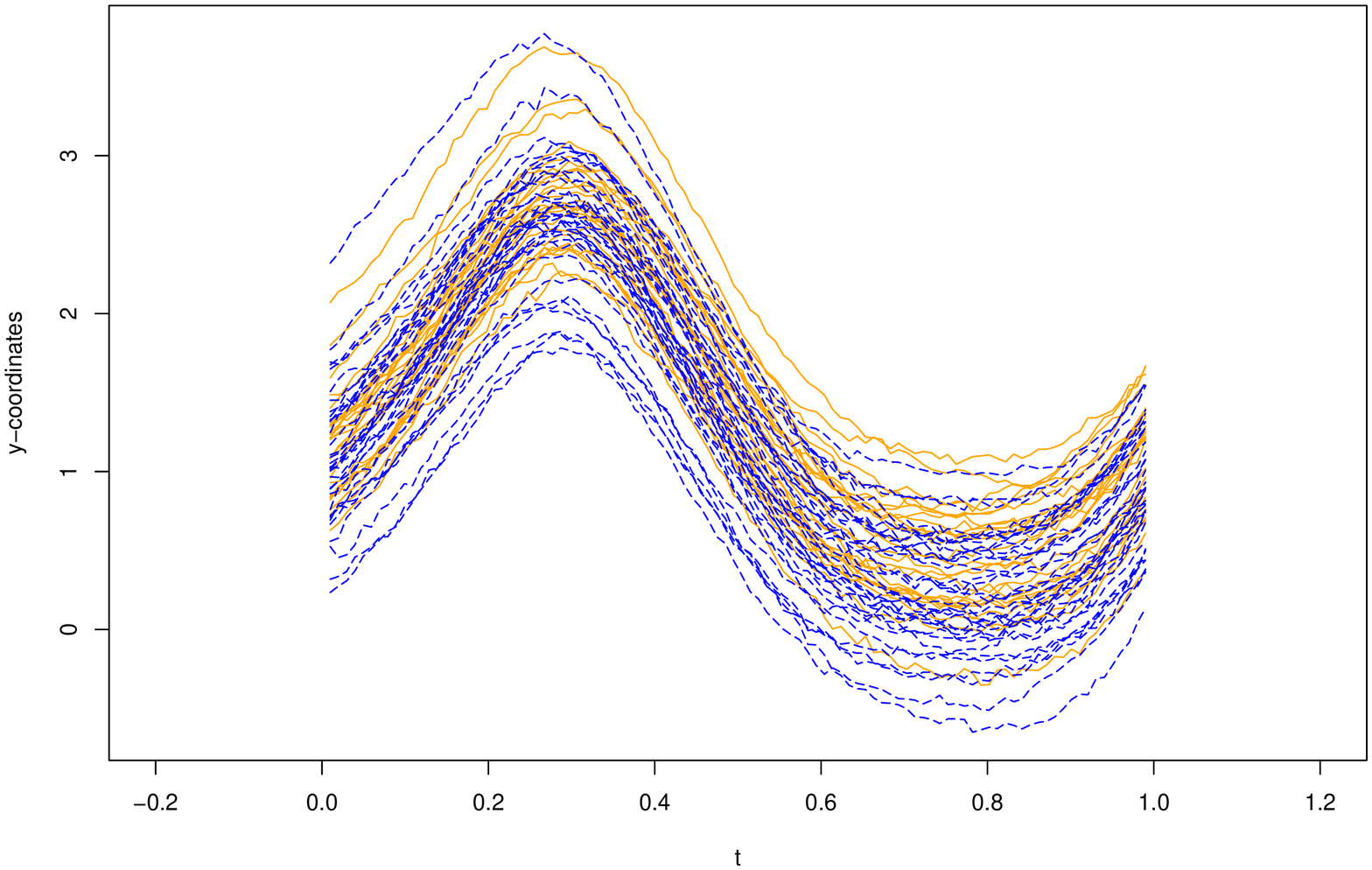}
		\caption{Aligned $\ve x_2(t)$}
	\end{subfigure}
	\quad

	\caption{An example of raw curves and corresponding aligned curves in Scenario B. Curves in solid orange lines represent the first group $(y=0)$ and the blue dashed lines represent the second group $(y=1)$. }
	\label{fig:5}       
\end{figure}

The simulation results were reported in Table~\ref{tab:3}. It can be obviously seen from Table~\ref{tab:3} that the proposed JCRC method performed quite well in prediction in the sense that all the values of CA, RI and ARI were relatively larger than those of other methods in both scenarios. On the other hand, as the degree of overlapping increased from Scenario A to Scenario B, the prediction accuracy decreased. Particularly, the SRV method performed worse among the others. Fig.\ref{fig:4} and Fig.\ref{fig:5} show the registration results by using JCRC method for Scenario A and Scenario B. The tuning parameters $K_x$ and $K_e$ in this simulation were selected to be $K_x=18$ and $K_e=10$ for Scenario A, $K_x=18$ and $K_e=12$ for Scenario B, respectively, by using a 5-fold Cross-Validation method.

\begin{table}[!htbp]
	\centering
	\caption{The performance of prediction between five different methods.}
	\label{tab:3}       
	\begin{tabular}{llllllllllllll}
		\hline\noalign{\smallskip}
		&&&\multicolumn{3}{c}{Scenario A}&&\multicolumn{3}{c}{Scenario B}\\
		\cline{4-6}\cline{8-10}
		\hspace{5mm}
		Methods&&&CA&RI&ARI&&CA&RI&ARI\\
		\hline

		JCRC   &&&$\ve{0.95}$&$\ve{0.90}$&$\ve{0.81}$    && $ \ve{0.94}$&$\ve{0.89}$&$\ve{0.79}$\\
		LLR      &&&0.85&0.78&0.49        &&   0.75&0.63&0.25\\
		JCRC-f &&&0.70&0.58&0.16      &&    0.78&0.66&0.32\\
		GPSM  &&&0.73&0.61&0.22      &&    0.76&0.64&0.27\\
		SRV     &&&0.56&0.52&0.03       &&    0.58&0.52&0.04\\
		\noalign{\smallskip}\hline
	\end{tabular}
\end{table}

\section{Real data example}
\label{sec:4}
\begin{figure}[!htbp]
	\centering
	\includegraphics[width=\textwidth,height=60mm]{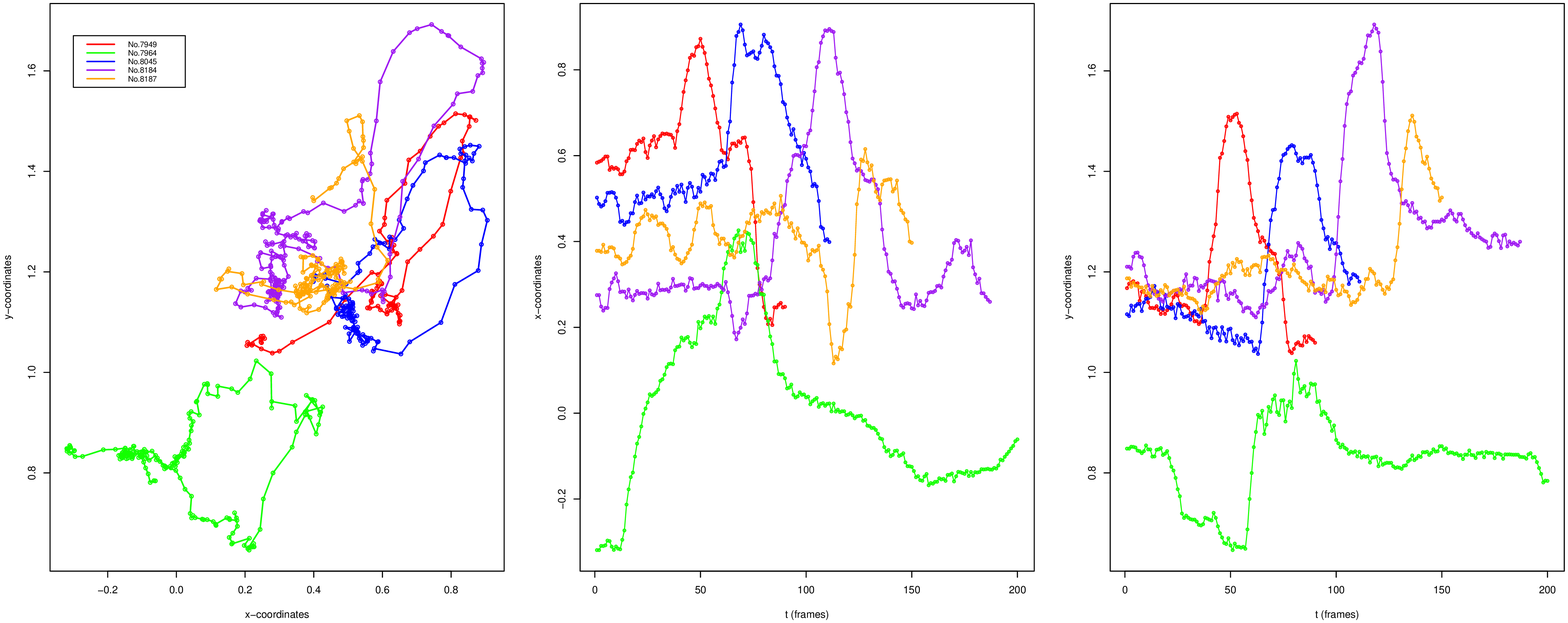}
	\caption{Trajectories of hyoid bone movement of five randomly selected subjects. The left panel represents the two-dimensional curves, the middle panel and the right panel illustrate the x-coordinates and the y-coordinates of the two-dimensional curves, respectively. }
	\label{fig:6}       
\end{figure}

\begin{figure}[!htbp]
	\centering
	\includegraphics[width=\textwidth]{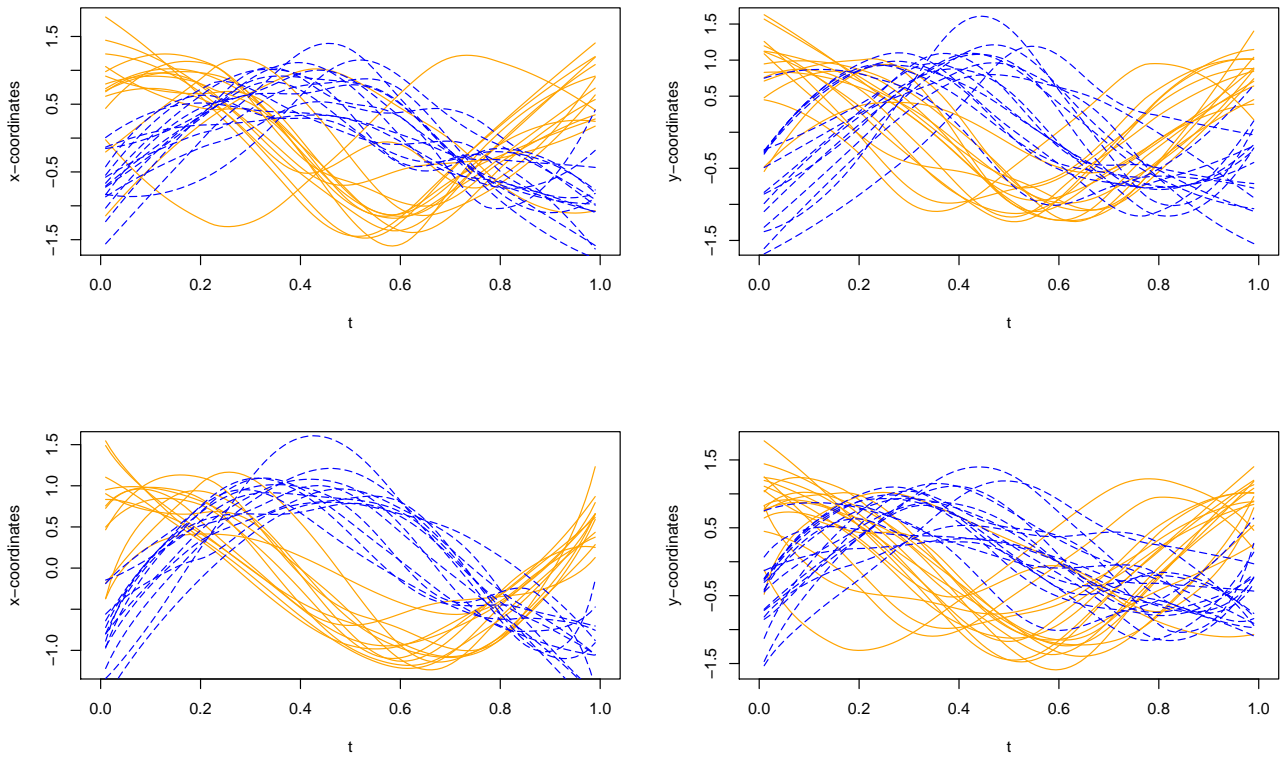}
	\caption{The raw curves (top panel) of x- and y-coordinates and their corresponding aligned curves (bottom panel) in the real data analysis. Curves in solid orange lines represent the first group $(y=0)$ and the blue dashed lines represent the second group $(y=1)$.}
	\label{fig:7}       
\end{figure}

\begin{figure}[!htbp]
	\centering
	\includegraphics[width=\textwidth,height=80mm]{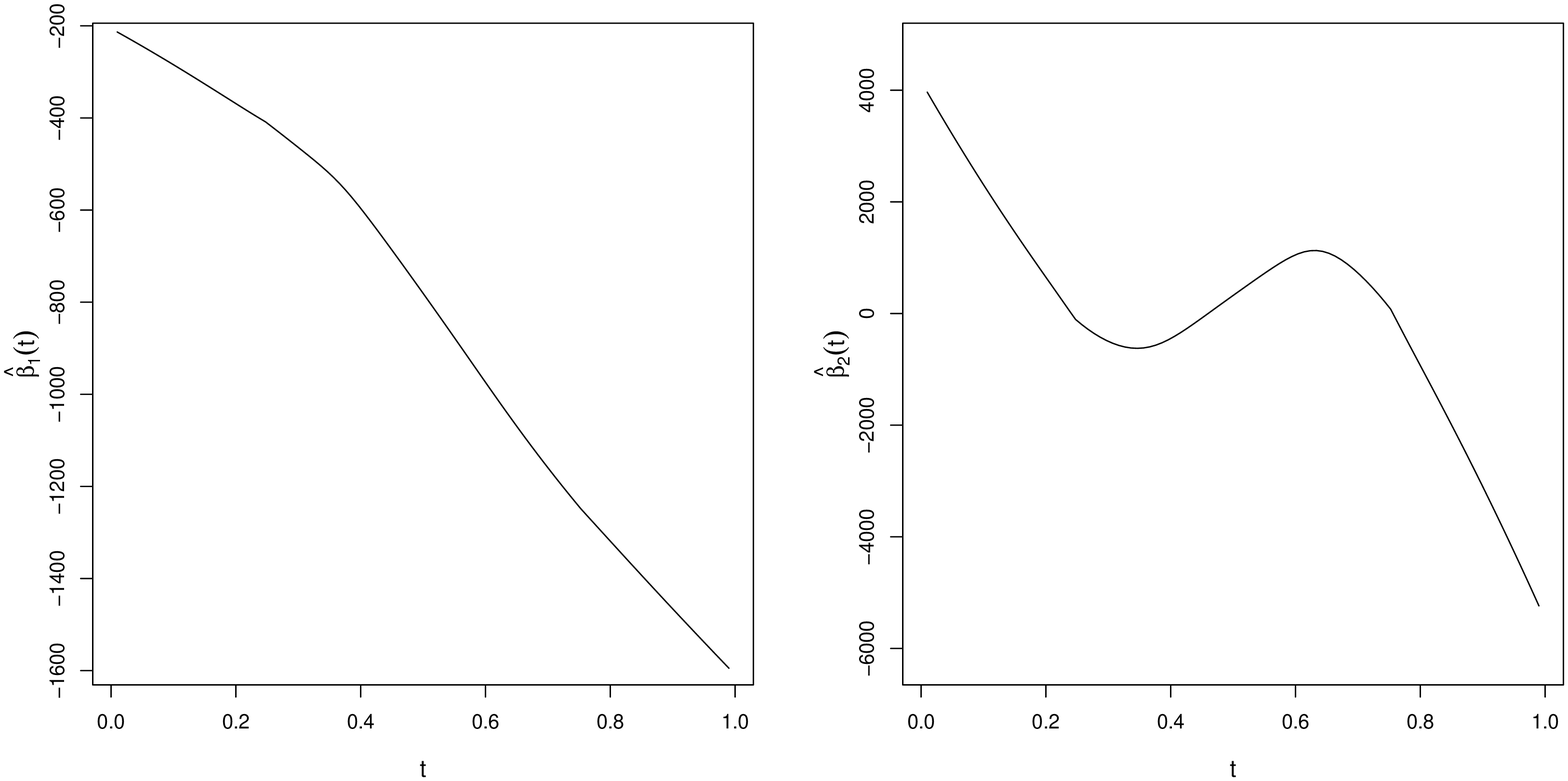}
	\caption{The estimation $\hat{\ve\beta}_1(t)$ (left panel) and $\hat{\ve\beta}_2(t)$ (right panel).}
	\label{fig:8}       
\end{figure}

The data we used to illustrate the proposed methodology are trajectories of the hyoid bone movement of stroke patients obtained from the database of videofluoroscopic swallow study (VFSS).
Patients after stroke usually suffer from oropharyngeal dysphagia, and the data obtained from the VFSS could be used to classify the dysphagia, to predict the prognosis or to assess the treatment effects \citep{kim2017}. A total of 30 subjects' data containing two groups, i.e. one for normal people and the other or patients after stroke, were obtained. Fig.\ref{fig:6} presented an example of the trajectories of hyoid bone for five subjects. We can see from Fig.\ref{fig:6} that there exists obvious misaligned problems for those curves in both vertical and horizontal variation. We fit the data using our proposed JCRC approach, in which the functional variables were taken to be the trajectories and the scalar variables were chosen to be motion time (duration), average velocity and acceleration amplitude of those curves. To evaluate the performance of classification, a 5-fold Cross-Validation method was used. We compared the performance of our proposed methodology with other four methods, and the results were reported in Table \ref{tab:4}. It can be seen from Table \ref{tab:4} that, the JCRC outperform other four methods in the sense that the values of CA, RI and ARI are relatively larger than those of other methods. Fig.\ref{fig:7} plotted the raw curves and the aligned curves in the real data analysis. The estimation $\hat{\beta}_1(t)$ and $\hat{\beta}_2(t)$ of the functional coefficients were shown in Fig.\ref{fig:8}.

\begin{table}[!htbp]
	\centering
	\caption{Classification results for the real data.}
	\label{tab:4}       
	\begin{tabular}{lllll}
		\hline\noalign{\smallskip}
		Methods&CA&RI&ARI\\
		\hline
		JCRC&$\ve{0.83}$&$\ve{0.69}$&$\ve{0.39}$\\
		LLR&0.63&0.45&0\\
		JCRC-f&0.43&0.45&0\\
		GPSM&0.57&0.48&0.06\\
		SRV&0.50&0.43&-0.11\\
		\noalign{\smallskip}\hline
	\end{tabular}
\end{table}

\section{Discussion}
\label{sec:5}
We proposed a joint model-based approach for simultaneously classifying and aligning the curves. A two-level model is developed, in which a nonlinear mixed effects model with time warping function is used as the first level model for the misaligned curves, providing simultaneous registration and modelling for the functional variables. A functional logistic regression model is then used as the second level model to predict the binary outcome representing the class label of the data. The model allows to use both functional and scalar variables, giving a much better performance than the models using functional or scalar variables only. Simulation studies and application to the hyoid bone movement data show that the proposed methodology work well in predicting the class label of the observed functional data. Under some regularity conditions, we investigated the identifiability of the data registration model and the asymptotic properties of the estimators as well.

In this study, a fast fitting method \citep{goldsmith2011} is applied in the estimation of the second level  model due to it's computational efficiency and it's good performance even $\ve{x}(\hat{g}^{-1}(t))$ is poorly obtained. Other alternatives, however, can be developed and it is also worth a further development on how to estimate the functional coefficients in the proposed model with sparsely observed noisy-corrupted curves. Moreover, some other sophisticated models will be considered for simultaneously classifying and aligning the misaligned curves, for example, Gaussian process priors in the models might be replaced by other heavy-tailed processes, which result in more robust models.

The R code is available upon request.

\bibliography{reference}

\begin{thebibliography}{}

\bibitem[Capra and M{\"u}ller, 1997]{capra1997}
Capra, W.~B. and M{\"u}ller, H.~G. (1997).
\newblock An accelerated-time model for response curves.
\newblock {\em J Am Stat Assoc}, 92:72--83.

\bibitem[Chamroukhi and Nguyen, 2019]{chamroukhi2019model}
Chamroukhi, F. and Nguyen, H.~D. (2019).
\newblock Model‐based clustering and classification of functional data.
\newblock {\em Wiley Interdisciplinary Reviews-Data Mining and Knowledge
  Discovery}, 9(4).

\bibitem[Delaigle and Hall, 2012]{delaigle2012achieving}
Delaigle, A. and Hall, P. (2012).
\newblock Achieving near perfect classification for functional data.
\newblock {\em J R Stat Soc B}, 74(2):267--286.

\bibitem[Gasser and Kneip, 1995]{gasser1995}
Gasser, T. and Kneip, A. (1995).
\newblock Searching for structure in curve samples.
\newblock {\em J Am Stat Assoc}, 90:1179--1188.

\bibitem[Gervini and Gasser, 2004]{gervini2004}
Gervini, D. and Gasser, T. (2004).
\newblock Self-modelling warping functions.
\newblock {\em J R Stat Soc B}, 66:959--971.

\bibitem[Goldsmith et~al., 2011]{goldsmith2011}
Goldsmith, J., Bobb, J., Crainiceanu, C.~M., Caffo, B., and Reich, D. (2011).
\newblock Penalized functional regression.
\newblock {\em J Comput Graph Stat}, 20:830--851.

\bibitem[Gower, 1975]{gower1975}
Gower, J.~C. (1975).
\newblock Generalized procrustes analysis.
\newblock {\em Psychometrika}, 40:33--51.

\bibitem[Hubert and Arabie, 1985]{hubert1985}
Hubert, L. and Arabie, P. (1985).
\newblock Comparing partitions.
\newblock {\em J Classif}, 2:193--218.

\bibitem[James, 2002]{james2002}
James, G.~M. (2002).
\newblock Generalized linear models with functional predictors.
\newblock {\em J R Stat Soc B}, 64:411--432.

\bibitem[James and Hastie, 2001]{james2001}
James, G.~M. and Hastie, T.~J. (2001).
\newblock Functional linear discriminant analysis for irregularly sampled
  curves.
\newblock {\em J R Stat Soc B}, 63:533--550.

\bibitem[Kim et~al., 2017]{kim2017}
Kim, W.-S., Zeng, P., Shi, J.~Q., Lee, Y., and Paik, N.-J. (2017).
\newblock Semi-automatic tracking, smoothing and segmentation of hyoid bone
  motion from videofluoroscopic swallowing study.
\newblock https://doi.org/10.1371/journal.pone.0188684.

\bibitem[Kneip et~al., 2000]{kneipetal2000}
Kneip, A., Li, X., MacGibbon, K.~B., and Ramsay, J.~O. (2000).
\newblock Curve registration by local regression.
\newblock {\em Canadian Journal of Statistics}, 28:19--29.

\bibitem[Li and Yu, 2008]{li2008}
Li, B. and Yu, Q. (2008).
\newblock Classification of functional data: A segmentation approach.
\newblock {\em Comput Stat Data Anal}, 52:4790--4800.

\bibitem[Liu and M{\"u}ller, 2004]{liu2004}
Liu, X. and M{\"u}ller, H.~G. (2004).
\newblock Functional convex averaging and synchronization for time-warped
  random curves.
\newblock {\em J Am Stat Assoc}, 99:687--699.

\bibitem[Liu and Yang, 2009]{liu2009simultaneous}
Liu, X. and Yang, M. C.~K. (2009).
\newblock Simultaneous curve registration and clustering for functional data.
\newblock {\em Comput Stat Data Anal}, 53(4):1361--1376.

\bibitem[Mallor et~al., 2018]{mallor2018}
Mallor, F., Moler, J.~A., and Urmeneta, H. (2018).
\newblock Simulation of household electricity consumption by using functional
  data analysis.
\newblock {\em Journal of Simulation}, 12:271--282.

\bibitem[McCulloch et~al., 2008]{mcculloch2008}
McCulloch, C.~E., Searle, C., and Neuhaus, J. (2008).
\newblock {\em Generalized, Linear, and Mixed Models}.
\newblock Wiley.

\bibitem[Mosler and Mozharovskyi, 2014]{mosler2014fast}
Mosler, K. and Mozharovskyi, P. (2014).
\newblock Fast dd-classification of functional data.
\newblock {\em arXiv: Methodology}.

\bibitem[M{\"u}ller, 2005]{mullerH2005}
M{\"u}ller, H. (2005).
\newblock Functional modelling and classification of longitudinal data.
\newblock {\em Scand J Stat}, 32(2):223--240.

\bibitem[M{\"u}ller and Stadtm{\"u}ller, 2005]{muller2005}
M{\"u}ller, H.-G. and Stadtm{\"u}ller, U. (2005).
\newblock Generalized functional linear models.
\newblock {\em Ann Stat}, 33:774--805.

\bibitem[Raket et~al., 2016]{raket2016separating}
Raket, L.~L., Grimme, B., Schoner, G., Igel, C., and Markussen, B. (2016).
\newblock Separating timing, movement conditions and individual differences in
  the analysis of human movement.
\newblock {\em PLOS Computational Biology}, 12(9).

\bibitem[Ramsay and Silverman, 2005]{ram05}
Ramsay, J.~O. and Silverman, B.~W. (2005).
\newblock {\em Functional Data Analysis}.
\newblock Springer-Verlag New York, USA.

\bibitem[Ramsey and Li, 1998]{ramsay1998}
Ramsey, J. and Li, X. (1998).
\newblock Curve registraion.
\newblock {\em J R Stat Soc B}, 60:351--363.

\bibitem[Rand, 1971]{rand1971}
Rand, W.~M. (1971).
\newblock Objective criteria for the evaluation of clustering methods.
\newblock {\em J Am Stat Assoc}, 66:846--850.

\bibitem[R{\o}nn, 2001]{ronn2001}
R{\o}nn, B.~B. (2001).
\newblock Nonparametric maximum likelihood estimation for shifted curves.
\newblock {\em J R Stat Soc B}, 63:243--259.

\bibitem[Ruppert, 2002]{ruppert2002}
Ruppert, D. (2002).
\newblock Selecting the number of knots for penalized splines.
\newblock {\em J Comput Graph Stat}, 11:735--757.

\bibitem[Sangalli et~al., 2010]{sangalli2010k-mean}
Sangalli, L.~M., Secchi, P., Vantini, S., and Vitelli, V. (2010).
\newblock k-mean alignment for curve clustering.
\newblock {\em Comput Stat Data Anal}, 54(5):1219--1233.

\bibitem[Shi et~al., 2012]{shi2012}
Shi, J.~Q., Wang, B., Will, E.~J., and West, R.~M. (2012).
\newblock Mixed‐effects gaussian process functional regression models with
  application to dose–response curve prediction.
\newblock {\em Stat Med}, 31:3165--3177.

\bibitem[Srivastava et~al., 2011]{srivastava2011}
Srivastava, A., Wu, W., Kurtek, S., Klassen, E., and Marron, J.~S. (2011).
\newblock Registration of functional data using fisher-rao metric.
\newblock https://arxiv.org/abs/1103.3817.

\bibitem[Ullah and Finch, 2013]{ullah2013}
Ullah, S. and Finch, C.~F. (2013).
\newblock Applications of functional data analysis: A systematic review.
\newblock {\em BMC medical research methodology}, 13:43.

\bibitem[Wang and Gasser, 1997]{wang1997}
Wang, K.~M. and Gasser, T. (1997).
\newblock Alignment of curves by dynamic time warping.
\newblock {\em Ann Stat}, 25:1251--1276.

\bibitem[Yu et~al., 2016]{yu2016}
Yu, D., Kong, L., and Mizera, I. (2016).
\newblock Partial functional linear quantile regression for neuroimaging data
  analysis.
\newblock {\em Neurocomputing}, 195:74--87.

\bibitem[Zeng et~al., 2019]{zeng2019}
Zeng, P., Shi, J.~Q., and Kim, W.-S. (2019).
\newblock Simultaneous registration and clustering for multidimensional
  functional data.
\newblock {\em J Comput Graph Stat}, 28(4):943--953.

\bibitem[Zhang and Telesca, 2014]{zhang2014joint}
Zhang, Y. and Telesca, D. (2014).
\newblock Joint clustering and registration of functional data.
\newblock arXiv: Methodology. https://arxiv.org/abs/1403.7134. Accessed 9
  August 2020.

\bibitem[Zhu et~al., 2012]{zhu2012}
Zhu, H., Brown, P.~J., and Morris, J.~S. (2012).
\newblock Robust classification of functional and quantitative image data using
  functional mixed models.
\newblock {\em Biometrics}, 68:1260--1268.

\end{thebibliography}


\begin{thebibliography}{}

\bibitem[Cai et~al., 2006]{cai2006}
Cai, T.~T., Hall, P., et~al. (2006).
\newblock Prediction in functional linear regression.
\newblock {\em Ann Stat}, 34:2159--2179.

\bibitem[Cao et~al., 2018]{cao2018}
Cao, R., Du, J., Zhou, J., and Xie, T. (2018).
\newblock Fpca-based estimation for generalized functional partially linear
  models.
\newblock {\em Statistical Papers}, pages 1--21.

\bibitem[Geyer, 1994]{geyer1994}
Geyer, C.~J. (1994).
\newblock On the convergence of monte carlo maximum likelihood calculations.
\newblock {\em J R Stat Soc B}, 56:261--274.

\bibitem[Hall et~al., 2007]{hall2007}
Hall, P., Horowitz, J.~L., et~al. (2007).
\newblock Methodology and convergence rates for functional linear regression.
\newblock {\em Ann Stat}, 35:70--91.

\bibitem[Li and Zhu, 2018]{li2018}
Li, T. and Zhu, Z. (2018).
\newblock Inference for generalized partial functional linear regression.
\newblock http://doi.org/10.5705/ss.202018.0155.

\bibitem[Pollard, 1991]{pollard1991}
Pollard, D. (1991).
\newblock Asymptotics for least absolute deviation regression estimators.
\newblock {\em Econometric Theory}, 7:186--199.

\bibitem[Shin, 2009]{shin2009}
Shin, H. (2009).
\newblock Partial functional linear regression.
\newblock {\em Journal of Statistical Planning and Inference}, 139:3405--3418.

\bibitem[Van~der Vaart, 2000]{van2000}
Van~der Vaart, A.~W. (2000).
\newblock {\em Asymptotic statistics}.
\newblock Cambridge university press.

\bibitem[Wang et~al., 2011]{wang2011}
Wang, L., Liu, X., Liang, H., and Carroll, R.~J. (2011).
\newblock Estimation and variable selection for generalized additive partial
  linear models.
\newblock {\em Ann Stat}, 39:1827.

\bibitem[White, 1996]{white1996}
White, H. (1996).
\newblock {\em Estimation, inference and specification analysis}.
\newblock Number~22. Cambridge university press.

\end{thebibliography}
\bibliographystyle{apalike}

\end{document}


\maketitle

\section*{Assumptions and technical proofs for asymptotic properties}
\paragraph{Proof of Theorem 1} The identifiability proof goes as follows. By assumptions in model (1), the random effects and the random error can be integrated into a new Gaussian process error denoted as $\epsilon^*_i(t)$ with $E\{\epsilon^*_i(t)\}=0$. Then we have that $\epsilon^*_i(t)=x_i(t)-\tau(g_i(t))$, thus there is no ambiguity about the error term. Suppose that $E\{x_i(t)\}=\tau_1(g_{1i}(t))=\tau_2(g_{2i}(t))$ for all $i=1,\ldots,N$, then we have $\tau_1(t)=\tau_2(g_{2i}(g^{-1}_{1i}(t)))$. Since the left-hand side of this equation doesn't depend on $i$, we have that $g_{2i}(g^{-1}_{1i}(t))=l(t)$ for all $i$ some function $l(\cdot)$. Then $g^{-1}_{1i}(t)=g^{-1}_{2i}(l(t))$ for all $i$. The assumption in Theorem 1 shows that $E(g_{1i})=E(g_{2i})$, then we have $l(t)=t$. Therefore, $g_{1i}(t)=g_{2i}(t)$ for all $i$ and the warping functions are identifiable.$\hfill\Box$

Let $\langle \cdot,\cdot\rangle$ and $\|\cdot\|$ be the $L^2(\cal T)$ inner product and norm, respectively. For notational simplicity, let $\tilde{\ve {x}}_{ai}(t)=\ve {x}_{ai}(g^{-1}(t))$ and define the covariance of the curve $\tilde{\ve {x}}_{ai}(t)$ as $K_{x_{a}}(s, t) = \text{Cov}[ \tilde{\ve {x}}_{ai}(s),  \tilde{\ve {x}}_{ai}(t)]$ , where $K_{x_{a}}(s,t)$ is continuous on the interval $[0,1]$. Then by Mercer's Theorem we have
\[
K_{x_{a}}(s, t)=\sum\limits_{l=1}^\infty\lambda_{al}\phi_{al}(s)\phi_{al}(t),
\]
where $\lambda_{a1}>\lambda_{a2}>\ldots>0$ are eigenvalues and $\phi_{a1}(t),\phi_{a2}(t)\ldots$ are eigenfunctions of the covariance operator corresponding to $K_{x_{a}}(s,t)$.
Then by the Karhunen-Lo$\grave{e}$ve representation, the process $\tilde{\ve x}_{ai}$ and the functional coefficient $\ve\beta_a(t)$ follows the following eigen decompositions
\[
\tilde{\ve x}_{ai}(t)=\sum\limits_{l=1}^\infty p_{ail}\phi_{al}(t) \hspace{5mm} \ve\beta_a(t)=\sum\limits_{l=1}^\infty e_{al}\phi_{al}(t),
\]
where $p_{ail}=\langle  \tilde{\ve x}_{ai}, \phi_{al}\rangle$ are uncorrelated random variables with zero mean and variance $\lambda_{al}$ and $e_{al}=\langle \ve \beta_a, \phi_{al}\rangle$.

In practice, $K_{x_{a}}(s,t)$ is unknown and we can take the empirical version by
$$\hat{K}_{x_{a}}(s,t)=\sum\limits_{l=1}^\infty\hat{\lambda}_{al}\hat{\phi}_{al}(s)\hat{\phi}_{al}(t),$$
where $(\hat{\lambda}_{al},\hat{\phi}_{al})$ is the estimator of $(\lambda_{al},\phi_{al})$ with $\hat{\lambda}_{a1}\geq\hat{\lambda}_{a2}\ldots\geq 0$.

Therefore, the systematic component in Equation~(12) can be rewritten as
\[
\eta_i =\ve v^{\T}_i\ve b+\int_0^1\tilde{\ve x}_{i}(t)\ve\beta(t)dt
= \ve v^{\T}_i\ve b+\sum\limits_{a=1}^2\sum\limits_{l=1}^\infty \tilde{p}_{ail}e_{al}
= \ve v^{\T}_i\ve b+\sum\limits_{a=1}^2\sum\limits_{l=1}^{K_x} \tilde{p}_{ail}e_{al}+R_i,
\]
where $\tilde{p}_{ail}=\langle  \tilde{\ve x}_{ai}, \hat{\phi}_{al}\rangle$, $R_i=\sum\limits_{a=1}^2\sum\limits_{l=K_x+1}^{\infty} \tilde{p}_{ail}e_{al}$ and $K_x$ is the tuning parameter which is set to be sufficiently large.

For notational simplicity, we define $$\tilde{\ve p}_i=(\tilde{\ve {p}}^{\T}_{1i}, \tilde{\ve {p}}^{\T}_{2i})^{\T}=(\tilde{p}_{1i1},\ldots,\tilde{p}_{1iK_{x}},\tilde{p}_{2i1},\ldots,\tilde{p}_{2iK_{x}})^{\T},$$
$$\ve {e}=(\ve {e}^{\T}_1, \ve {e}^{\T}_2)^{\T}=(e_{11},\ldots,e_{1K_x},e_{21},\ldots,e_{2K_x})^{\T},$$ 
$$\ve {Y}=(y_1,\ldots,y_N)^{\T},$$ 
$$\ve {v}=(\ve {v}^{\T}_1,\ldots,\ve {v}^{\T}_N)^{\T},$$
$$\tilde{\ve {X}}=(\tilde{\ve {x}}_1(t),\ldots,\tilde{\ve {x}}_N(t)),$$  $$\ve{\eta}=(\eta_1,\ldots,\eta_N)^{\T}.$$

Let $\ve {U}=(\ve {v}, \tilde{\ve {X}})$ and $\ve{\theta}=(\ve {b}^{\T},\ve {e}^{\T})^{\T}$ be the unknown parameter vector of the models defined in Equation~(11) and Equation (12). Then the estimation $\hat{\ve{\theta}}=(\hat{\ve {b}}^{\T},\hat{\ve {e}}^{\T})^{\T}$ is obtained by maximizing the following log-likelihood function
\[
\ell(\ve{\theta})\overset{\triangle}{=}\ell(\ve {Y}; \ve{\eta})
=\sum\limits_{i=1}^N\ell_i(y_i; \eta_i)
=\sum\limits_{i=1}^N\big\{y_i\ve{\zeta}(\ve v_i,\ve {x}_i)+{\cal B}[\ve\zeta(\ve{v}_i,\ve {x}_i)]+{\cal C}(y_i)\big\}
\]

For simplicity, let $C$ be a constant whose value might change according to different circumstances. Denote $\dot{\ell}_{i,d}(y_i;d)$ and $\ddot{\ell}_{i,d}(y_i;d)$ as the first- and second-order derivative of $\ell_i(y_i; d)$ with respect to $d$, respectively. Also, similar to \cite{li2018}, we define
\[
\text{I}(\ve U)=-\text{E}\big[\ddot{\ell}_{\eta}(\ve Y; \eta)|\ve U\big]=-\text{E}\big[\sum\limits_{i=1}^N\ddot{\ell}_{i,\eta_{i}}(y_i; \eta_i)|\ve U\big],
\]
where $\eta_i=\ve v^{\T}_i\ve b+\int_{0}^{1}\tilde{\ve x}_i(t)\ve\beta(t)dt$.

In model (11), the response variable $y_i$ is related to both the scalar variables and the functional variables. So, the main complicated issue comes from the dependence between $\ve {v}_i$ and $\tilde{\ve {x}}_i(t)$. To solve this problem, similar to \cite{shin2009}, we define
\[
\ve {G}(\tilde{\ve {X}})=\frac{\text{E}(\ve {v}\ddot{\ell}_{\eta}(\ve {Y}; \eta)|\tilde{\ve {X}})}{\text{E}(\ddot{\ell}_{\eta}(\ve {Y}; \eta)|\tilde{\ve {X}})},~\text{and}~\ve {v}=\tilde{\ve {v}}+\ve {G}(\tilde{\ve {X}}),
\]
where $\tilde{\ve {v}}=(\tilde{v}_1,\ldots,\tilde{v}_{p+1})^{p+1}$ is a zero mean $(p+1)$-dimensional random vector and $\ve {G}(\tilde{\ve {X}})=(G_1(\tilde{\ve {X}}),\ldots,G_{p+1}(\tilde{\ve {X}}))^{\T}$ is a $(p+1)$-dimensional functional vector with $G_j(\tilde{\ve{X}})\in L^2(\cal T)$ for $j=1,\ldots,p+1$.

Suppose that the following assumptions hold

\begin{description}[]\label{assumptions}
	\item[(S1).] For each $i\in \{1,\ldots,N\}$, $\text{E}\|\tilde{\ve {x}}_i\|^4<\infty$. \label{assum1}
	
	\item [(S2).] For each $a\in[1,2]$ and $l$, $\text{E}(p^4_{ail})\leq C\lambda^2_{al}$ with the eigenvalue $\lambda_{al}$ satisfies $C^{-1}l^{-\alpha}\leq \lambda_{al}\leq Cl^{-\alpha}$ and $\lambda_l-\lambda_{a(l+1)}\geq Cl^{-\alpha-1}$ for $l\geq 1$ and some constant $\alpha>1$. In addition, $|e^*_{al}|\leq Cl^{-\gamma}$ for some constant $\gamma>\alpha/2+1$.
	\label{assum2}
	
	\item[(S3).] The tuning parameter $K_x$ satisfies
	$$K_x\asymp N^{\frac{1}{\alpha+2\gamma}},$$
	where the notation $a_N\asymp b_N$ means that there exist constants $0<L<M<\infty$ such that $L\leq a_N/b_N\leq M$ for all N. 
	\label{assum3}
	
	\item[(S4).] For each $i$, the scalar covariates $\ve {v}_i$ satisfies $\text{E}\|\ve {v}_i\|^4<\infty$.
	\label{assum4}
	
	\item[(S5).] $\text{E}(\tilde{\ve {v}})=\ve {0}$,
	$$\ve{\Omega}_1=\text{E}\big\{\sum\limits_{i=1}^N\ddot{\ell}_{i,\eta^*_{i}}(y_i; \eta^*_{i})\tilde{\ve {v}}_i\tilde{\ve {v}}^{\T}_i\big\}~\text{and}~\ve{\Omega}_2=\text{E}\big\{\sum\limits_{i=1}^N\dot{\ell}^2_{i,\eta^*_{i}}(y_i; \eta^*_{i})\tilde{\ve {v}}_i\tilde{\ve {v}}^{\T}_i\big\},$$ where $\ve{\Omega}_1$ and $\ve{\Omega}_2$ are assumed to be positive definite matrices and $\eta^*_i=\ve {v}^{\T}_i\ve {b}^*+\int_0^1\tilde{\ve {x}}_i(t)\ve{\beta}^*(t)$.
	\label{assum5}
	
	\item[(S6).] $|I(\ve {U})|<C$ and $I(\ve {U})$ satisfies the first-order Lipschitz condition.
	\label{assum6}
	
	\item [(S7).]The true value $\ve{\theta}^*_x$ of $\ve{\theta}_x$ is unique and $\hat{\ve{\theta}}_x\overset{p}{\rightarrow} \ve{\theta}^*_x$ where $\hat{\ve{\theta}}_x$ is the MLE of $\ve{\theta}_x$.
	\label{assum7}
	
	\item [(S8).]For $i=1,\ldots,N$, the likelihood function $\ell_i(\ve{\theta}_x)$ of the second level model is thrice continuously differentiable with respect to $\ve{\theta}_x$.
	\label{assum8}
	
	\item [(S9).]There exist positive definite matrices $\ve {A}(\ve\theta^*_x)$ and $\ve {B}(\ve\theta^*_x)$ such that
	\[
	\underset{N\rightarrow\infty}{\text{lim}}-\frac{1}{N}\sum\limits_{i=1}^N\partial^2_{\theta_x}\ell_i(\ve{\theta}^*_x)=\ve {A}(\ve{\theta}^*_x),~ \underset{N\rightarrow\infty}{\text{lim}}\frac{1}{N}\sum\limits_{i=1}^N\partial_{\theta_x}\ell_i(\ve{\theta}^*_x)\partial_{\theta_x}\ell_i(\ve{\theta}^*_x)^{\T}=\ve {B}(\ve{\theta}^*_x).
	\]
	\label{assum9}
\end{description}

\begin{remark}
	Assumptions  \ref{assum1}-\ref{assum3} are required similar to the settings in the functional linear models \citep{cai2006,hall2007} while assumptions \ref{assum4}-\ref{assum6} are used to deal with the linear part with scalar  variables. Assumptions \ref{assum7}-\ref{assum9} are used to develop the asymptotical consistency of the MLE for the first-level model.
\end{remark}

In order to develop the proofs of of the theorems, the following lemmas are provided. Let  $\tilde{\eta}_i=\ve {v}^{\T}_i\ve {b}^*+\tilde{\ve {p}}^{\T}_i\ve {e}^*$, where $\ve b^*$ and $\ve e^*$ denote the true values of $\ve b$ and $\ve e$, respectively.

\begin{lemma}\label{lem1}
	Let $R_i=\int_0^1\tilde{\ve x}_i(t)\ve\beta^*(t)dt-\tilde{\ve p}^{\T}_i\ve e^*$ for $i=1,\ldots,N$. Then under assumptions \ref{assum1}-\ref{assum6}, we have
	$$\|R_i\|^2=\|\eta^*_{i}-\tilde{\eta}_i\|^2=O_p(N^{-(2\gamma+\alpha-1)/(\alpha+2\gamma)}).$$
	Thus
	$$\eta^*_{i}-\tilde{\eta}_i=O_p(N^{-(2\gamma+\alpha-1)/2(\alpha+2\gamma)}).$$
\end{lemma}

\paragraph{Proof}
	Note that 
	\begin{eqnarray*}
		R_i&=&\int_{0}^{1}\tilde{\ve x}_i(t)\ve\beta^*(t)dt-\tilde{\ve p}^{\T}_i\ve e^*\\
		&=&\sum\limits_{a=1}^2\sum\limits_{l=1}^\infty\langle \tilde{\ve x}_{ai}, \phi_{al}\rangle\langle\ve\beta^*_a, \phi_{al}\rangle-\sum\limits_{a=1}^2\sum\limits_{l=1}^{K_x}\langle \tilde{\ve x}_{ai}, \hat{\phi}_{al}\rangle\langle\ve\beta^*_a, \phi_{al}\rangle\\
		&=&\sum\limits_{a=1}^2\sum\limits_{l=1}^{K_x}\langle \tilde{\ve x}_{ai}, \phi_{al}\rangle\langle\ve\beta^*_a, \phi_{al}\rangle-\sum\limits_{a=1}^2\sum\limits_{l=1}^{K_x}\langle \tilde{\ve x}_{ai}, \hat{\phi}_{aj}\rangle\langle\ve\beta^*_a, \phi_{al}\rangle\\
		&+&\sum\limits_{a=1}^2\sum\limits_{l=K_x+1}^\infty\langle \tilde{\ve x}_{ai}, \phi_{al}\rangle\langle\ve\beta^*_a, \phi_{al}\rangle\\
		&=&\sum\limits_{a=1}^2\sum\limits_{l=1}^{K_x}\langle \tilde{\ve x}_{ai}, \phi_{al}- \hat{\phi}_{al}\rangle e^*_{al}+\sum\limits_{a=1}^2\sum\limits_{l=K_x+1}^\infty \langle \tilde{\ve x}_{ai}, \phi_{al}\rangle e^*_{al}\\
		&=&I_1+I_2
	\end{eqnarray*}
where $I_1=\sum\limits_{a=1}^2\sum\limits_{l=1}^{K_x}\langle \tilde{\ve x}_{ai},  \phi_{al}- \hat{\phi}_{al}\rangle e^*_{al}$, and\\ $I_2=\sum\limits_{a=1}^2\sum\limits_{l=K_x+1}^\infty \langle \tilde{\ve x}_{ai}, \phi_{al}\rangle e^*_{al}=\sum\limits_{a=1}^2\sum\limits_{l=K_x+1}^\infty p_{ail} e^*_{al}$.

Since $\|\hat{\phi}_{al}-\phi_{al}\|^2=O_p(N^{-1}l^2)$, then it follows from assumptions \ref{assum1}-\ref{assum3} that
\begin{eqnarray*}
	\|I_1\|^2&=&\|\sum\limits_{a=1}^2\sum\limits_{l=1}^{K_x}\langle \tilde{\ve x}_{ai}, \phi_{al}- \hat{\phi}_{al}\rangle e^*_{al}\|^2\\
	&\leq& 2K_x\sum\limits_{l=1}^{K_x}\|\langle \tilde{\ve x}_{ai}, \phi_{al}- \hat{\phi}_{al}\rangle e^*_{al}\|^2\\
	&\leq &2K_x\sum\limits_{l=1}^{K_x}\|\phi_{al}- \hat{\phi}_{al}\|^2|e^*_{al}|^2\\
	&\leq& O_P(K_x)\sum\limits_{l=1}^{K_x}N^{-1}l^2l^{-2\gamma}\\
	& \leq & O_P(N^{-1}K_x)\sum\limits_{l=1}^{K_x}l^{-2\gamma+2}\\
	& \leq & O_P(N^{-1}K_x)=O_P(N^{-\frac{2\gamma+\alpha+1}{\alpha+2\gamma}})\\
\end{eqnarray*}

For $I_2$, note that $p_{ail}$ are uncorrelated random variables with zero mean and variance $\lambda_{al}$, then we have
\begin{eqnarray*}
	\text{E}(\|I_2\|^2)&=&\text{E}(\sum\limits_{a=1}^2\sum\limits_{l=K_x+1}^\infty p_{ail} e^*_{al})^2\\
	&=&  \sum\limits_{a=1}^2\sum\limits_{l=K_x+1}^\infty e^{*2}_{al}\lambda_{al}\\
	&\leq &2C\sum\limits_{l=K_x+1}^\infty l^{-(2\gamma+\alpha)}\\
	&=&O_p(K^{-(2\gamma+\alpha-1)}_x)=O_p(N^{-\frac{2\gamma+\alpha-1}{\alpha+2\gamma}})\\
\end{eqnarray*}

Then 
$R_i=I_1+I_2=O_p(N^{-\frac{2\gamma+\alpha-1}{2(\alpha+2\gamma)}})+O_p(N^{-\frac{2\gamma+\alpha-1}{2(\alpha+2\gamma)}})=O_p(N^{-\frac{2\gamma+\alpha-1}{2(\alpha+2\gamma)}})$ holds,
which indicates that $\|R_i\|^2=\|\eta^*_{i}-\tilde{\eta}_i\|^2=O_p(N^{-(2\gamma+\alpha-1)/(\alpha+2\gamma)})$, thus Lemma \ref{lem1} holds. $\hfill\Box$

Let $\tilde{\ve v}_i=\ve v_i-\ve G(\tilde{\ve x}_i)$ and 
\[
\breve{\ve b}=\underset{\ve b}{\text{arg max}}\sum\limits_{i=1}^N\ell_i(y_i; \tilde{\ve v}^{\T}_i\ve b+\tilde{\ve p}^{\T}_i\ve e^*+\ve G(\tilde{\ve x}_i)\ve b^*).
\]

Then the following Lemma says that the estimation $\breve{\ve b}$ is asymptotically distributed as normal distribution.
\begin{lemma}\label{lem2}
	Under assumptions \ref{assum1}-\ref{assum9}, we have
	\[
	\sqrt{N}(\breve{\ve b}-\ve b^*)\rightarrow \text{N}(\ve 0, \ve\Omega^{-1}_1\ve\Omega_2\ve\Omega^{-1}_1),
	\]
	where  $\ve\Omega_1=\text{E}\big\{\sum\limits_{i=1}^N\ddot{\ell}_{i,\eta^*_{i}}(y_i; \eta^*_{i})\tilde{\ve v}_i\tilde{\ve v}^{\T}_i\big\}$, $\ve\Omega_2=\text{E}\big\{\sum\limits_{i=1}^N\dot{\ell}^2_{i,\eta^*_{i}}(y_i; \eta^*_{i})\tilde{\ve v}_i\tilde{\ve v}^{\T}_i\big\}$ and $\tilde{\ve v}_i=\ve v_i-\ve G(\tilde{\ve x}_i)$.
\end{lemma}

\paragraph{Proof}
Let $\ve \omega=\sqrt{N}(\ve b-\ve b^*)$ and $\breve{\ve \omega}=\sqrt{N}(\breve{\ve b}-\ve b^*)$, which according to the definition of $\breve{\ve b}$, is obtained by maximizing the following function
\[
M(\ve \omega)=\sum\limits_{i=1}^N\ell_i(y_i; \tilde{\ve v}^{\T}_i\ve b^*+\tilde{\ve p}^{\T}_i\ve e^*+\ve G(\tilde{\ve x}_i)\ve b^*+\tilde{\ve v}^{\T}_i\ve \omega/\sqrt{N})-\sum\limits_{i=1}^N\ell_i(y_i; \ve v^{\T}_i\ve b^*+\tilde{\ve p}^{\T}_i\ve e^*).\]

Taking a second-order Taylor's expansion of $M(\ve\omega)$ yields
$$M(\ve\omega)=\frac{1}{\sqrt{N}}\sum\limits_{i=1}^N\dot{\ell}_{i,\tilde{\eta}_i}(y_i;\tilde{\eta}_i)\tilde{\ve v}^{\T}_i\ve\omega+\frac{1}{2}\ve\omega^{\T}\ve\Sigma \ve\omega,$$
where
$$\ve\Sigma=\frac{1}{N}\sum\limits_{i=1}^N\ddot{\ell}_{i,\tilde{\eta}_i}(y_i; \tilde{\eta}_i+\nu_i)\tilde{\ve v}_i\tilde{\ve v}^{\T}_i$$
with $\nu_i$ lies between 0 and $\tilde{\ve v}_i\ve\omega/\sqrt{N}$. It follows from \cite{van2000} that $\ve\Sigma=-\ve \Omega_1+o_p(1)$.

On the other hand,
\begin{align*}\label{taylor}
\frac{1}{N}\sum\limits_{i=1}^N\dot{\ell}_{i,\tilde{\eta}_i}(y_i;\tilde{\eta}_i)\tilde{\ve v}^{\T}_i&=\frac{1}{\sqrt{N}}\sum\limits_{i=1}^N\dot{\ell}_{i,\eta^*_{i}}(y_i;\eta^*_{i})\tilde{\ve v}^{\T}_i\ve+\frac{1}{\sqrt{N}}\sum\limits_{i=1}^N\ddot{\ell}_{i,\eta^*_{i}}(y_i;\eta^*_{i})\tilde{\ve v}^{\T}_i(\tilde{\eta_{i}}-\eta^*_{i})\\
&+o_p(1)\\
&=I_3+I_4+o_p(1)\\
\end{align*}
where $$ I_3=\frac{1}{\sqrt{N}}\sum\limits_{i=1}^N\dot{\ell}_{i,\eta^*_{i}}(y_i;\eta^*_{i})\tilde{\ve v}^{\T}_i\ve~\text{and}~ I_4=\frac{1}{\sqrt{N}}\sum\limits_{i=1}^N\ddot{\ell}_{i,\eta^*_{i}}(y_i;\eta^*_{i})\tilde{\ve v}^{\T}_i(\tilde{\eta_{i}}-\eta^*_{i}).$$

It is easy to find that $I_4=o_p(1)$. By the Lindeberg-Feller central limit theory, we have $I_3\rightarrow\text{N}(\ve 0, \ve\Omega_2)$. Then, 
$$M(\ve\omega)= I^{\T}_3\ve\omega-\frac{1}{2}\ve\omega^{\T}\ve\Omega_1 \ve\omega+o_p(1).$$

The results of \cite{geyer1994} and \cite{pollard1991} show that
$$\breve{\ve\omega}=\ve \Omega^{-1}_1 I_3+o_p(1),$$ 
then Lemma \ref{lem2} holds from the Slutsky Theorem.$\hfill\Box$

\begin{lemma}\label{lem3}
	Under assumptions \ref{assum1}-\ref{assum9}, we have
	$$\|\hat{\ve\theta}-\breve{\ve\theta}\|^2=O_p(N^{-\frac{2\gamma-1}{\alpha+2\gamma}}),$$
	where $\hat{\ve\theta}=(\hat{\ve b}^{\T},\hat{\ve e}^{\T})^{\T}$ and $\breve{\ve\theta}=(\breve{\ve b}^{\T}, \ve e^{*\T})^{\T}$.
\end{lemma}

\paragraph{Proof}
Taking a first-order Taylor's expansion of $\dot{\ell}(\hat{\ve\theta})=\frac{\partial\ell(\theta)}{\partial\theta}|_{\theta=\hat{\theta}}$ at $\breve{\ve\theta}$ yields
$$0=\dot{\ell}(\hat{\ve\theta})=\dot{\ell}(\breve{\ve\theta})+\ddot{\ell}(\bar{\ve\theta})(\hat{\ve\theta}-\breve{\ve\theta})+o_p(1),$$
where $\bar{\ve\theta}$ lies between $\hat{\ve\theta}$ and $\breve{\ve\theta}$, $\dot{\ell}(\breve{\ve\theta})=\frac{\partial\ell(\theta)}{\partial\theta}|_{\theta=\breve{\theta}}$ and $\dot{\ell}(\bar{\ve\theta})=\frac{\partial\ell(\theta)}{\partial\theta}|_{\theta=\bar{\theta}}$.

Then we have
$$\hat{\ve\theta}-\breve{\ve\theta}=-[\ddot{\ell}(\bar{\ve\theta})]^{-1}\dot{\ell}(\breve{\ve\theta}).$$

Denote
\[
\dot{\ell}(\breve{\ve\theta})=\bigg\{(\frac{\partial\ell(\breve{\ve\theta})}{\partial\ve b})^{\T}, (\frac{\partial\ell(\breve{\ve\theta})}{\partial\ve e})^{\T}\bigg\}^{\T}\\
=\sum\limits_{i=1}^N\dot{\ell}_{i,\breve{\eta}_i}(y_i; \breve{\eta}_i)(\ve v^{\T}_i,\tilde{\ve p}^{\T}_i)^{\T},
\]
where $\breve{\eta}_i=\ve v^{\T}_i\breve{\ve b}+\tilde{\ve p}^{\T}_i\ve e^*$.

Note that
$$\frac{\partial\ell(\breve{\ve\theta})}{\partial\ve b}=\sum\limits_{i=1}^N\dot{\ell}_{i,\breve{\eta}_i}(y_i; \breve{\eta}_i)\ve v_i=\sum\limits_{i=1}^N\dot{\ell}_{i,\eta^*_{i}}(y_i; \eta^*_{i})\ve v_i+\sum\limits_{i=1}^N\ddot{\ell}_{i,\bar{\eta}_i}(y_i; \bar{\eta}_i)(\ve v^{\T}_i(\breve{\ve b}-\ve b^*)+R_i)\ve v_i,$$
and
$$\frac{\partial\ell(\breve{\ve\theta})}{\partial\ve e}=\sum\limits_{i=1}^N\dot{\ell}_{i,\breve{\eta}_i}(y_i; \breve{\eta}_i)\tilde{\ve p}_i=\sum\limits_{i=1}^N\dot{\ell}_{i,\eta^*_{i}}(y_i; \eta^*_{i})\tilde{\ve p}_i+\sum\limits_{i=1}^N\ddot{\ell}_{i,\bar{\eta}_i}(y_i; \bar{\eta}_i)(\tilde{\ve p}^{\T}_i(\breve{\ve e}-\ve e^*)+R_i)\tilde{\ve p}_i,$$
where $\bar{\eta}_i$ lies between $\eta^*_{i}$ and $\breve{\eta}_i$.

Similar to \cite{cao2018}, we have
\begin{equation}\label{routine}
\text{E}\big(\|\sum\limits_{i=1}^N\dot{\ell}_{i,\eta^*_{i}}(y_i; \eta^*_{i})\ve v_i\|\big)=O(N^{1/2}).
\end{equation}

The assumption \ref{assum2} and Lemma \ref{lem1} indicate that
\begin{equation}\label{approx}
\begin{array}{llll}
\|\sum\limits_{i=1}^N\ddot{\ell}_{i,\bar{\eta}_i}(y_i; \bar{\eta}_i)(\ve v^{\T}_i(\breve{\ve b}-\ve b^*)+R_i)\ve v_i\|&=&O_p(N^{1/2})+O_p(N\cdot N^{-(2\gamma+\alpha-1)/2(\alpha+2\gamma)})\\
&=&O_p(\sqrt{NK_x}).\\
\end{array}
\end{equation}

Equation (\ref{routine}) and (\ref{approx}) show that $\partial\ell(\breve{\ve\theta})/\partial\ve b=O_p(\sqrt{NK_x})$. Similarly, we have $\partial\ell(\check{\ve\theta})/\partial\ve e=O_p(\sqrt{NK_x})$. Therefore, $\dot{\ell}(\breve{\ve\theta})=O_p(\sqrt{NK_x})$.

Similar to Lemma A.3 of \cite{wang2011}, we have $\big\|\big(\frac{1}{N}\ddot{\ell}(\bar{\ve\theta})\big)^{-1}\big\|=O_p(\lambda^{-1/2}_{K_x})=O_p(K^{\alpha/2}_x)$, which yields
\[
\begin{array}{lll}
\|\hat{\ve\theta}-\breve{\ve\theta}\|&\leq&\big\|\big(\frac{1}{N}\ddot{\ell}(\bar{\ve\theta})\big)^{-1}\big\|\big\|\frac{1}{N}\dot{\ell}(\breve{\ve\theta})\big\|\\
&=&O_p(K^{\alpha/2}_x)O_p(N^{-(2\gamma+\alpha-1)/2(\alpha+2\gamma)})\\
&=&O_p(N^{-(2\gamma-1)/2(\alpha+2\gamma)})\\
\end{array}
\]
Thus, the result $\|\hat{\ve\theta}-\breve{\ve\theta}\|^2=O_p(N^{-(2\gamma-1)/(\alpha+2\gamma)})$ holds. $\hfill\Box$

The proofs for the Theorems and the Corollary are presented as follows.
\paragraph{Proof of Theorem 2}
Let $\hat{\eta}_i=\ve v^{\T}_i\hat{\ve b}+\tilde{\ve p}^{\T}_i\hat{\ve e}$ and for any $\ve z\in\mathbb {R}^{p+1}$, define $\hat{\eta}_i(\ve z)=\ve v^{\T}_i\hat{\ve b}+\tilde{\ve p}^{\T}_i\hat{\ve e}+\tilde{\ve v}^{\T}_i\ve z$, where $\tilde{\ve v}_i=\ve v_i-\ve G(\tilde{\ve x}_i)$.  Obviously, when $\ve z=\ve 0$,  
$$\hat{\eta}(\ve z)=\underset{\eta(\ve z)}{\text{arg max}}\ell(\ve Y;  \eta(\ve z)).$$

Then the following equation follows from a Taylor's expansion
\[
\begin{array}{lll}
0=\frac{\partial\ell(\hat{\eta}(\ve z))}{\partial\ve z}|_{\ve z=\ve 0}&=&\sum\limits_{i=1}^N\dot{\ell}_{i,\hat{\eta}_i}(y_i; \hat{\eta}_i)\tilde{\ve v}_i
\\
&=&\sum\limits_{i=1}^N\dot{\ell}_{i,\eta^*_{i}}(y_i; \eta^*_{i})\tilde{\ve v}_i+\sum\limits_{i=1}^N\ddot{\ell}_{i,\eta^*_{i}}(y_i; \eta^*_{i})\tilde{\ve v}_i(\hat{\eta}_i-\eta^*_{i})+o_p(1),\\
\end{array}
\]

Applying Lemma~\ref{lem2} and Lemma~\ref{lem3}, the second term on the right hand side of the above equation can be rewritten as
$$\frac{1}{N}\sum\limits_{i=1}^N\ddot{\ell}_{i,\eta^*_{i}}(y_i; \eta^*_{i})\tilde{\ve v}_i(\hat{\eta}_i-\eta^*_{i})=\frac{1}{N}\sum\limits_{i=1}^N\ddot{\ell}_{i,\eta^*_{i}}(y_i; \eta^*_{i})\tilde{\ve v}_i\tilde{\ve v}^{\T}_i(\hat{\ve b}-\ve b^*)+o_p(N^{1/2}).$$
Then we have 
$$(\hat{\ve b}-\ve b^*)=-\big[\frac{1}{N}\sum\limits_{i=1}^N\ddot{\ell}_{i,\eta^*_{i}}(y_i; \eta^*_{i})\tilde{\ve v}_i\tilde{\ve v}^{\T}_i\big]^{-1}\big[\frac{1}{N}\sum\limits_{i=1}^N\dot{\ell}_{i,\eta^*_{i}}(y_i; \eta^*_{i})\tilde{\ve v}_i\big]+o_p(N^{-1/2}).$$

Therefore, by Central Limit Theory and Slutsky's Theorem, we have
$$\sqrt{N}(\hat{\ve b}-\ve b^*)\rightarrow \text{N}(\ve 0, \ve\Omega^{-1}_1\ve\Omega_2\ve\Omega^{-1}_1),$$
where $\ve\Omega_1$ and $\ve\Omega_2$ are defined in Lemma~\ref{lem2}.$\hfill\Box$

\paragraph{Proof of Theorem 3}
Similar to \cite{shin2009}, for any $a\in[1,2]$, we have
\begin{align*}
\|\hat{\ve\beta}_a(t)-\ve\beta^*_{a}\|&=\big\|\sum\limits_{l=1}^{K_x}\hat{e}_{al}\hat{\phi}_{al}-\sum\limits_{l=1}^\infty e^*_{al}\phi_{al}\big\|^2\\
&\leq 2\big\|\sum\limits_{l=1}^{K_x}\hat{e}_{al}\hat{\phi}_{al}-\sum\limits_{l=1}^{K_x} e^*_{al}\phi_{al}\big\|^2+2\big\|\sum\limits_{l=K_x+1}^\infty e^*_{al}\phi_{al}\big\|^2\\
&\leq 4\big\|\sum\limits_{l=1}^{K_x}(\hat{e}_{al}-e^*_{al})\hat{\phi}_{al}\big\|^2+4\big\|\sum\limits_{l=1}^{K_x}e^*_{al}(\hat{\phi}_{al}-\phi_{al})\big\|^2+2\sum\limits_{l=K_x+1}^\infty e^{*2}_{al}\\
&\leq 4\|\hat{\ve e}_{a}-\ve e^*_{a}\|^2+8K_x\sum\limits_{l=1}^{K_x}e^{*2}_{al}\|\hat{\phi}_{al}-\phi_{al}\|^2+2\sum\limits_{l=K_x+1}^\infty e^{*2}_{al}.\\
\end{align*}

Note that
\begin{equation}\label{term1}
\sum\limits_{l=K_x+1}^\infty e^{*2}_{al}\leq C\sum\limits_{l=K_x+1}^\infty l^{-2\gamma}=O(K^{-(2\gamma-1)}_x)=O(N^{-(2\gamma-1)/(\alpha+2\gamma)}),
\end{equation}
and by $\|\hat{\phi}_{al}-\phi_{al}\|^2=O_p(N^{-1}l^2)$, we have
\begin{equation}\label{term2}
8K_x\sum\limits_{l=1}^{K_x}e^{*2}_{al}\|\hat{\phi}_{aj}-\phi_{aj}\|^2\leq O_p(N^{-1}K_x)=o_p(N^{-(2\gamma-1)/(\alpha+2\gamma)}).
\end{equation}

In addition, from Lemma~\ref{lem3} we have $\|\hat{\ve e}-\ve e^*\|^2=O_p(N^{-(2\gamma-1)/(\alpha+2\gamma)})$, then Theorem 3 holds by combining this with equation (\ref{term1}) and (\ref{term2}).$\hfill\Box$

\paragraph{Proof of Corollary 1}
Let $\hat{\eta}_i=\ve v^{\T}_i\hat{\ve b}+\int_0^1\tilde{\ve x}_i(t)\hat{\ve\beta}(t)dt$, where $\hat{\ve b}$ and $\hat{\ve\beta}(t)$ are obtained from our proposed estimation procedure. Then, we have
\begin{align*}
\hat{\eta}_i-\eta^*_i &= \ve v^{\T}_i\hat{\ve b}+\int_0^1\tilde{\ve x}_i(t)\hat{\ve\beta}(t)dt-\big[\ve v^{\T}_i\ve b^*+\int_0^1\tilde{\ve x}_i(t)\ve\beta^*(t)dt\big]\\
&=\ve v^{\T}_i(\hat{\ve b}-\ve b^*)+\sum\limits_{a=1}^2\sum\limits_{l=1}^{\infty}\langle \tilde{\ve x}_{ai},\phi_{al}\rangle\langle\hat{\ve\beta}_a,\hat{\phi}_{al}\rangle-\sum\limits_{a=1}^2\sum\limits_{l=1}^{\infty}\langle \tilde{\ve x}_{ai},\phi_{al}\rangle\langle\ve\beta^*_a,\phi_{al}\rangle\\
&=\ve v^{\T}_i(\hat{\ve b}-\ve b^*)+\sum\limits_{a=1}^2\sum\limits_{l=1}^{\infty}\langle \tilde{\ve x}_{ai},\phi_{al}\rangle(\hat{e}_{al}-e^*_{al})\\
&=I_5+I_6,\\
\end{align*}
where $\hat{e}_{al}=\langle\hat{\ve\beta}_a,\hat{\phi}_{al}\rangle$, $e^*_{al}=\langle\ve\beta^*_a,\phi_{al}\rangle$, $I_5=\ve v^{\T}_i(\hat{\ve b}-\ve b^*)$ and \\
$I_6=\sum\limits_{a=1}^2\sum\limits_{l=1}^{\infty}\langle \tilde{\ve x}_{ai},\phi_{al}\rangle(\hat{e}_{al}-e^*_{al})=\sum\limits_{a=1}^2\sum\limits_{l=1}^{\infty}p_{ail}(\hat{e}_{al}-e^*_{al})$.

Given Theorem 2, the fact that $\|\hat{\ve b}-\ve b^*\|=O_p(N^{-1/2})$ holds. Then under assumption \ref{assum4} we have $\text{E}\|I_5\|^2=\|\hat{\ve b}-\ve b^*\|^2\text{E}\ve v^2_i=O_p(N^{-1})$.

Since $\langle \tilde{\ve x}_{ai},\phi_{al}\rangle$ are uncorrelated random variables with zero mean and variance $\lambda_{al}$, and by Lemma 3 we have $\|\hat{\ve e}-\ve e^*\|^2=O_p(N^{-(2\gamma-1)/(\alpha+2\gamma)})$. Then under assumption \ref{assum2}, we have
\[
\text{E}(\|I_6\|^2) 
=\sum\limits_{a=1}^2\sum\limits_{l=1}^\infty\lambda_{al}\|\hat{e}_{al}-e^*_{al}\|^2
\leq 2O_p(N^{\frac{2\gamma-1}{\alpha+2\gamma}})C\sum\limits_{l=1}^\infty l^{-\alpha}
=O_p(N^{\frac{2\gamma-1}{\alpha+2\gamma}}).
\]
Thus
\[
\hat{\eta}_i-\eta^*=I_5+I_6=O_p(N^{-1/2})+O_p(N^{-(2\gamma-1)/(\alpha+2\gamma)})=O_p(N^{-1/2}).
\]
Note that the functional logistic regression model is a special case of (11) and (12) with a logistic link function, i.e. $\eta_i=\rm{logit}(\pi_i)$. Since the inverse link function $h^{-1}(\eta_i)$ is continuous and differentiable in $\eta_i$, then $h^{-1}(\hat{\eta}_i)-h^{-1}(\eta^*_i)=O_p(N^{-1/2})$ holds, which indicate that $\hat{\pi}_i-\pi^*_i=O_p(N^{-1/2})$.$\hfill\Box$

\paragraph{Proof of Theorem 4}
$\hat{\ve\theta}_x$ is the MLE of the second level model obtained through conditional models described in Web Appendix A, then under assumptions \ref{assum7}-\ref{assum9}, the theorem follows from \cite{white1996} immediately, to save space, we omit the proof here.$\hfill\Box$

\bibliography{reference}
\bibliographystyle{apalike}